\documentclass[12pt,preprint]{aastex}

\newcommand{\cind}[1]{\noindent{\bf #1\/:} }
\newcommand{\cosec}{\mathop{\rm cosec}\nolimits}
\newcommand{\merr}[3]{$#1_{#2}^{#3}$}

\slugcomment{AJ, in press}

\shorttitle{MS4 Sample.  II}
\shortauthors{Burgess \& Hunstead}

\begin{document}

\title{The Molonglo Southern 4\,Jy Sample (MS4).  II. \\
    ATCA Imaging and Optical Identification}

\author{A. M. Burgess\altaffilmark{1}}
\email{annb@psych.usyd.edu.au}

\and

\author{R. W. Hunstead}
\affil{School of Physics, University of Sydney,
    NSW 2006, Australia}
\email{rwh@physics.usyd.edu.au}

\altaffiltext{1}{present address: School of Psychology, University of
Sydney, NSW 2006, Australia.}

\clearpage

\begin{abstract}
Of the 228 sources in the Molonglo Southern 4\,Jy Sample (MS4), the 133
with angular sizes $<35''$ have been imaged at 5\,GHz at $2-4''$
resolution with the Australia Telescope Compact Array.  More than 90\%
of the sample has been reliably optically identified, either on the
plates of the UK Schmidt Southern Sky Survey or on $R$-band CCD images
made with the Anglo-Australian Telescope.  A subsample of 137 sources,
the SMS4, defined to be a close southern equivalent of the northern
3CRR sample, was found to have global properties mostly consistent with
the northern sample.  Linear sizes of MS4 galaxies and quasars were
found to be consistent with galaxy-quasar unification models of
orientation and evolution.
\end{abstract}


~~~~~~
\newpage

\section{INTRODUCTION}
\label{sec.intro}

In the preceding paper (\citealt{paper1}, hereafter Paper~I), the
Molonglo Southern 4\,Jy sample (MS4) was defined according to the
following selection criteria:  declination $-85^{\circ} < \delta <
-30^{\circ}$; flux density S$_{\rm 408\,MHz} > 4.0$\,Jy; Galactic
latitude $|b| > 10^{\circ}$; not in the Magellanic Cloud regions;
and not a known Galactic source.
The sample thus defined contained 228 radio sources.  These sources
were all imaged at 843\,MHz with the Molonglo Observatory Synthesis
Telescope (MOST) at a resolution of  $43'' \times 43'' \cosec
|\delta|$, giving radio positions accurate to around 1$-$2\arcsec, and
estimates of angular size.

A subsample of 137 sources, the SMS4, was defined to have S$_{\rm
178\,MHz} > 10.9$\,Jy, making it a southern equivalent of the northern
3CRR sample \citep{lrl}.  This sample was found to have properties mostly
consistent with those of 3CRR, though with slight, possibly significant
differences in angular-size distributions and source densities, in the
sense that SMS4 has slightly higher source density and larger angular
sizes than 3CRR.

In this paper we present higher-resolution radio images of those MS4
sources of smaller angular size (Section~\ref{section.atca}), as well
as optical identifications based on the Schmidt plates and on $R$-band
CCD images (Section~\ref{optid.ms4}).  Section~\ref{sec.indcomm}
contains notes on individual sources.  In Section~\ref{sec.sampcont} we
use the radio and optical data to compare derived properties of MS4,
SMS4, and 3CRR, such as their radio luminosity distributions, quasar
fractions, and radio linear sizes.

\section{RADIO IMAGING AT 5\,GHZ}
\label{section.atca}

We used our 843\,MHz MOST data from Paper~I to select the 133 sources
with largest angular size (LAS) $\leq 35''$ for high-resolution imaging
with the Australia Telescope Compact Array (ATCA).  The ATCA is an
east-west array consisting of six 22\,m-diameter paraboloidal antennas
\citep{main.AT.frater}.  We used an observing frequency of 5\,GHz, with
a bandwidth of 128\,MHz.  The observing was done in twelve different
runs between 1990 and 1993, at an early stage of ATCA operations.  The
first eight observing runs used five antennas with a longest spacing of
3\,km, with nominal resolution $\sim 4'' \times 4'' \cosec|\delta|$.
The last five observing runs were made after the 6\,km antenna was
commissioned, giving a resolution of $\sim 2'' \times 2'' \cosec
|\delta|$.

The observations were done in `cuts' or `snapshot' mode, a method of
time-sharing in which sources are observed at several hour angles
\citep{amb.snapshot}.  We usually observed about ten sources in a
twelve-hour run.  About eight cuts of 8-minutes' duration, spread in
hour angles, were observed per source, bringing the total integration
time per source to about one hour.  We included at least two secondary
phase calibrators in each cycle, with `cuts' of length four minutes,
and observed the primary flux-density calibrator, PKS~B1934$-$63, once
in each twelve-hour run.  Table~\ref{tab1.atobs} contains a summary of the
ATCA imaging runs.

A few sources were excluded either because they were unresolved ATCA
calibrators \\
(MRC~B1740$-$517 and MRC~B2259$-$375), or because they had
been imaged by others: \\
MRC~B0521$-$365 \citep{map.keel}, MRC~B0743$-$673
\citep{map.sproats,0743.rayner}, MRC~B1814$-$519
(M.~Wieringa, private communication), and MRC~B2153$-$699
\citep{map.firstat}.  MRC~B1933$-$587 had already been observed in a
different project.  We also observed a few sources which were more
extended than our 35\arcsec\ limit.  A summary of the observations for
each source is given in Table~\ref{tab2.atobs}.  The columns of this table
are as follows:

\begin{enumerate}

\item Source name.

\item Run number (see column~1 of Table~\ref{tab1.atobs}).

\item Number of cuts.

\item Total integration time in minutes.

\item Dynamic range of the final image, expressed as the ratio of the
peak flux density to the RMS noise.

\item FWHM major and minor axes of the CLEAN beam in arcseconds,
and position angle of the major axis in degrees east of north
(modulo 90).

\end{enumerate}

For most of our target sources, the ATCA snapshot observations enabled us
to obtain images with dynamic ranges of around 100:1.  These images provided
morphological information with a resolution of a few arcseconds, allowing
us to classify the radio structures and direct our searches for optical
identifications.

The data were reduced in the AIPS (Astronomical Image Processing
System) software package.  Because most of our observing was done before
polarisation calibration was feasible, we imaged only the total intensity.
Images were constructed using uniform weighting of the antenna baselines.
The procedures for deconvolution and self-calibration of ATCA snapshot
data are described in more detail in \citet{amb.snapshot}.

\subsection{Calibration and Imaging}

In continuum ATCA observing the 128\,MHz bandwidth is divided into 32
channels.  We usually averaged the central 20 channels, giving an
effective bandwidth of $\sim 80$\,MHz.  Because the ratio of bandwidth
to observing frequency was low, and the sources had angular extent $<
2'$ and were close to the field centre, any resultant bandwidth smearing
would not have been significant.

The flux-density scale was based on the radio spectrum of PKS~B1934$-$63
measured by \citet{reynolds.scale}.  Secondary phase calibrators were
observed at regular intervals to monitor the relative antenna gains.
A few secondary calibrators were found to be slightly resolved: \\
MRC~B0438$-$436, MRC~B1424$-$418, MRC~B1814$-$637, and MRC~B1827$-$360.
In these cases we either used the other secondary calibrator, or excluded
the baselines on which the calibrator was resolved.

\begin{figure}[p]
\includegraphics[bb=0 0 643 774,clip,height=627pt,width=442pt]{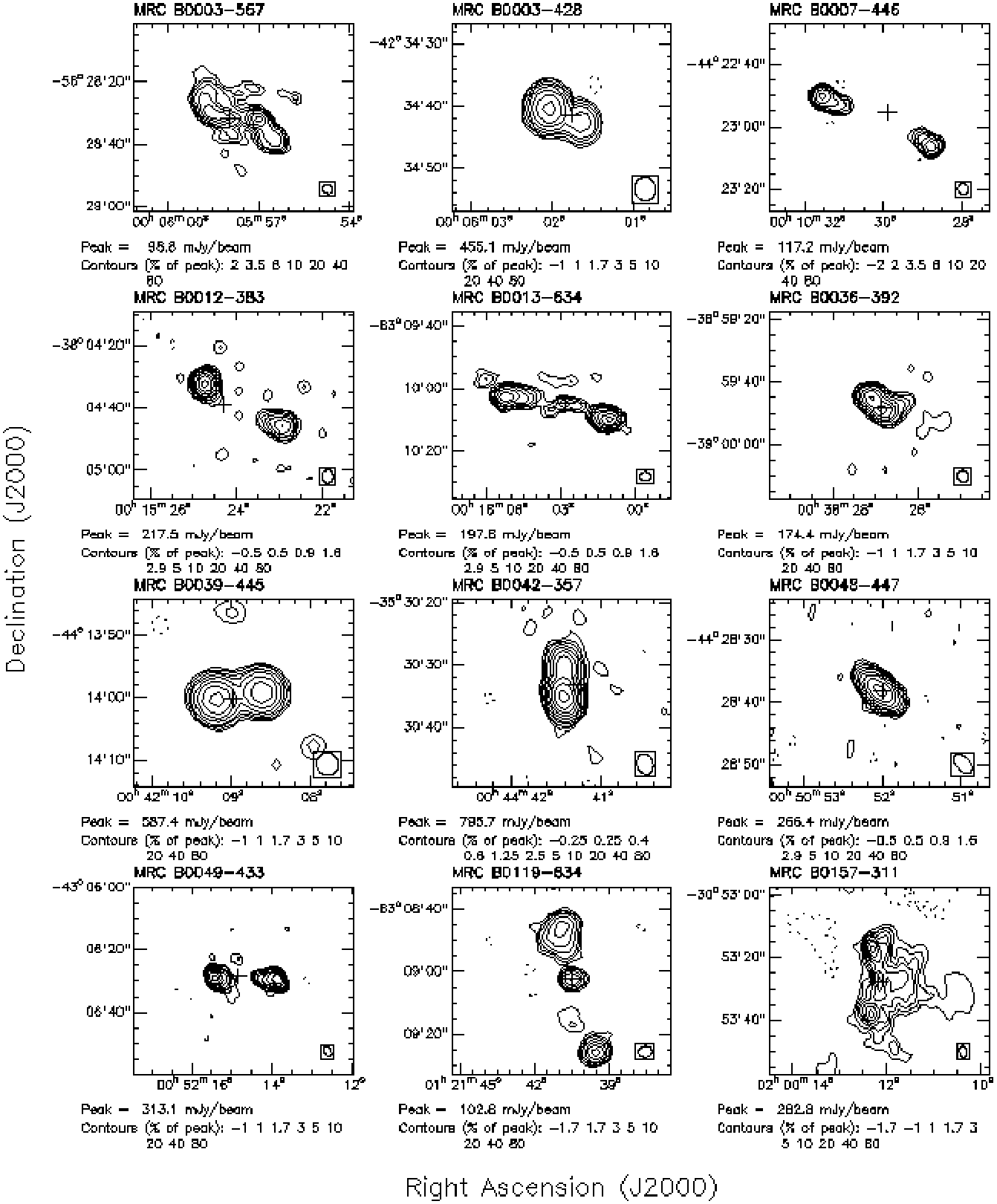}
\caption{ATCA contour maps\label{fig1.at}}
\end{figure}

The raw images were deconvolved with the Cotton-Schwab CLEAN algorithm
\citep{cotton.schwab}, using the AIPS task {\sc mx}.  The CLEANed images
typically contained artefacts caused by antenna phase gain variations
too rapid to be corrected by the secondary phase calibrator, limiting the
dynamic range to about 20:1.  Because most of the target sources were
strong (peak $S_{\rm 5\,GHz} > 100$mJy/beam) and compact (LAS $<$
35\arcsec), it was usually feasible to use self-calibration
\citep{asp6.selfcal} to improve the dynamic range to about 100:1.
Self-calibrating the phases was sufficient:  the strongest 200--300
CLEAN components were used as the starting model, which was updated
after each cycle.  We stopped self-calibrating when the dynamic range
of the image ceased to improve, usually after four to ten cycles.  The
final self-calibrated data set was then imaged and CLEANed, usually
with a loop gain of 0.1 and about 500 iterations.  Contour maps of
these final images are shown in Figure~\ref{fig1.at}, with crosses
denoting the positions of the likely optical counterpart.  Maps of
sources without resolved structure have been omitted.

We tested the reliability of our ATCA images by comparing them with Very
Large Array (VLA) images for 12 sources.  Structures for each source
showed good agreement above the 1--2\% contour levels.  A potential
hazard of self-calibration is the loss of absolute positional
information (e.g. \citealp{asp6.strat}).  These effects have been
reduced by applying successive models for self-calibration to the
original $uv$ data set.  Comparison with VLBI and optical data
(\S~\ref{sec.atpos}) shows that our positions are reliable at the
levels stated.

\subsection{Measurements from the ATCA Images}
\label{sec.atmeas}

Quantities measured from the radio images are summarised in
Table~\ref{tab3.atdat}.  The columns of the table are as follows:

\begin{enumerate}

\item Source name.

\item Code for Fanaroff-Riley \citep{fr} or other structural
classification.

\item Right ascension, in J2000 co-ordinates.

\item Declination, in J2000 co-ordinates.

\item Code for the type of position measurement (\S~\ref{sec.atpos}).

\item Integrated flux density in mJy (\S~\ref{sec.atflux}).  The
integrated flux densities are uncertain and sometimes underestimated,
due to the limited $uv$ coverage.  They have been measured mainly to
provide an estimate of the fraction of emission present in compact
components.

\item Core flux density in mJy.  In the absence of spectral-index
information, a compact component is called the `core' if it coincides
with an optical counterpart.  However, for close doubles identified
with galaxies, a component coincident with the optical counterpart has
not been classified as the core, because D2 structures \citep{d2.ref}
are rare in galaxies.  For most sources the core flux density is an
upper limit.  If extended emission lies near the core, the flux density
has been flagged with a colon (:).  For doubles with widely separated
lobes, we adopted an upper limit of three times the r.m.s. noise in a
box of size $6 \times 6$ pixels centred at the optical counterpart.  For
close doubles we set an arbitrary upper limit of $1/3$ of the integrated
flux density of the brighter lobe.  For compact-steep-spectrum (CSS)
sources no attempt was made to estimate the core flux density.

\item Ratio $R$ of core to extended flux density at 5\,GHz (observed).
These ratios are uncertain, because of the uncertainty in the core flux
densities.  $R$ was calculated from the core flux density in column 7
and the Parkes single-dish flux density (see Table~5 of Paper~I), which
was interpolated to the ATCA observing frequency using the radio
spectrum.  In a few cases the ATCA core flux density was greater than
the single-dish total flux density, presumably because of variability.
For these sources the ATCA integrated flux density was used instead of
the Parkes value.  Because of their uncertainty, the values of $R$ have
not been used elsewhere in the paper.

\item Largest angular extent in arcseconds (\S~\ref{sec.at.ang}).

\item Position angle of the extension in degrees east of north
(modulo 90).

\end{enumerate}

\subsection{Positions}
\label{sec.atpos}

Positions of compact sources (angular size $\lesssim 3''$) were measured
in AIPS with {\sc imfit}, which performs a least-squares fit to an elliptical
Gaussian.  Centroid positions of more extended sources were measured
above a contour level which depended on the dynamic range, but was
usually about 2\% of the peak.  For double sources with edge-brightened
(Fanaroff-Riley class II or FR\,2) structure, we also measured the peak
positions of the radio lobes so as to calculate the largest angular
extent.

To test the accuracy of the ATCA positions, the core positions of eight
quasars were compared with VLBI positions from \citet{vlbi.ma98}.
Because of the comparatively low resolution of the ATCA images, the
ATCA `core' position may be affected by blending.  This will increase
the scatter in the ATCA-VLBI offsets, so that the errors quoted below
may be slightly overestimated.  The mean offset in RA was 0\arcsec.07,
with $\sigma = 0''.32$.  The mean offset in declination was
$0''.08\cosec|\delta|$, with $\sigma = 0''.40\cosec|\delta|$.  The ATCA
position errors are therefore around 0\arcsec.3 in right ascension and
$0''.4\cosec|\delta|$ in declination.  The errors will be larger for
extended sources, and for a few sources with calibration problems.

Positions for sources in Run~9 (see Table~\ref{tab1.atobs}) were
affected by a telescope error and were therefore corrected with a
mean radio-optical offset in right ascension of 0\arcsec.9 ($\sigma =
1''.5$), and in declination of 1\arcsec.1$\cosec|\delta|$ ($\sigma =
0''.9\cosec|\delta|$), calculated from 6 sources with small angular
extent or a detected core.  The positions for Run~9 are flagged in
Table~\ref{tab3.atdat}.

\subsection{Flux Densities}
\label{sec.atflux}

Flux densities were measured mainly to estimate the fraction of
emission coming from compact components.  For compact sources we used
the integrated flux density of the Gaussian fitted with {\sc imfit}.
For extended sources of irregular shape we used the AIPS verb {\sc
tvstat}, which sums the flux density of each pixel in a polygonal
region specified by the user.  We also used {\sc tvstat} to correct for
a nonzero base level by finding the average flux density per pixel in
an annulus surrounding the source.

For slightly extended sources we measured integrated flux densities with
the method of \citet{rad.J1}, i.e. by calculating a set of volumes in
the radio image with upper bound given by the surface defined by the
radio contours, and lower bounds given by a set of lower flux-density
limits, and then extrapolating the results to the local zero level.

\subsection{Measurement of Angular Sizes and Position Angles}
\label{sec.at.ang}

For sources with a single compact component, the quoted angular size is
the deconvolved FWHM of the Gaussian fitted with {\sc imfit}.  Sources
with deconvolved angular sizes below a limit of $0.46$ times the major
axis of the dirty beam (a value chosen after examination of the
interferometer fringe visibilities for compact sources) were given this
value as an upper limit.

For edge-brightened (FR\,2) doubles, angular sizes and position angles
were defined using the separation between the outer peaks of the lobes,
usually the hot-spots.  For edge-darkened (FR\,1) doubles, and sources
of irregular structure, we measured the largest separation between
$3\sigma$ contours on the contour maps.  The position angle was taken
to be that of the line joining the two points which had the largest
separation.

Angular sizes were compared with those in Paper~I obtained with MOST.
For FR\,2 sources the ATCA angular size tends to be similar to the MOST
value, but slightly larger.   This is expected, both because the hot-spots
at the edge of the structure are resolved better with ATCA, and because
the steep-spectrum extended emission between the lobes contributes a
larger fraction of the flux density at 843\,MHz, reducing the width
of the fitted Gaussian.  For two FR\,1 sources, MRC~B1416$-$493 and
MRC~B2354$-$350, the ATCA detected only the inner structure, leading
to an underestimate of angular size.  The remaining FR\,1 sources had
larger angular sizes with ATCA than with MOST, owing to a difference in
definition: with MOST we measured the FWHM of an elliptical Gaussian,
whereas with the ATCA we measured the separation between outer contour
levels in a higher-resolution image.

\section{OPTICAL IDENTIFICATION OF THE SAMPLE}
\label{optid.ms4}

Optical identification of MS4 sources was conducted using first the UK
Schmidt sky survey, followed by $R$-band CCD imaging of the fainter
objects and blank fields at the prime focus of the Anglo-Australian
Telescope (AAT).  We used the identifications of \citet{rad.J1} for
objects which we had not imaged with MOST, as their optical positions
were measured in the same way as for the other MS4 sources.

Using Schmidt plates and AAT CCD images we found optical candidates for
222 of the 228 sources in the sample.  There are spectroscopic
redshifts in the literature or from our unpublished observations for
111 of the sources, mostly quasars and low-redshift galaxies.  Most of
the remaining optical counterparts are sufficiently faint that
measurement of spectroscopic redshifts will require a major observing
program on 4-m or 8-m telescopes.

\subsection{Measurement from the UK Schmidt Plates}
\label{schmi.meas}

Optical positions on film copies of the IIIaJ plates were
measured by B.~Piestrzynska using a purpose-built measuring machine
\citep{posflux.molonglo,opt.h11}.  These positions, hereafter referred
to as BP positions, are accurate to about 0\arcsec.5 in each co-ordinate
(\S~\ref{sec.opt.acc}).  Candidates were classified as galaxies or
stellar by visual inspection of the Schmidt plates.  We used the MOST
and ATCA positions to choose between candidates, using the following
search criteria:

\begin{enumerate}

\item For the 64 sources with LAS$\leq 8''$, the ATCA peak or the MOST
centroid from Paper~I was used as the search position.  In a few cases of
asymmetric structure in the ATCA image, the MOST centroid was preferred.

\item 75 sources had LAS$>$8\arcsec\ and a radio core position from the
ATCA image (\S~\ref{sec.atmeas}) or the literature.  For these the
radio core position was used as the search position.

\item 73 sources had LAS$>$8\arcsec\ and FR\,2 structure with no detected
radio core.  The search position was defined to be the ``midway point'':
the point midway between the MOST centroid and the midpoint of the
outermost hot spots in the ATCA image.  The justification for this choice
is described below.

\item 6 sources had LAS$>8''$ and FR\,1 structure with no measured radio
core position.  If there were two well defined inner peaks, their
midpoint was used as the search position, otherwise the MOST centroid
was used.

\item The MOST centroid was used for 10 sources with LAS$>8''$ with no
radio core position, and for which structure had not been classified as
FR\,1 or 2.

\end{enumerate}

For extended sources we used a circular search area of radius 0.2 times
the largest angular extent, centred at the search position.  For point
sources and radio cores we checked if the displacement vector for the
radio-optical offset fell within an elliptical search area with semiminor
axis proportional to the radio and optical right ascension errors summed
in quadrature, and semimajor axis proportional to the radio and optical
declination errors summed in quadrature.  For sources with LAS $<8''$ the
constant of proportionality used was 2, and for radio cores it was 2.5.
The search area had to be slightly larger for the radio cores, because
core positions were often affected by extended emission in the image.

The search position for FR\,2 sources was chosen empirically to be the
point midway between the radio centroid and the midpoint of the radio
hot-spots.  In various studies, the radio midpoint (e.g.
\citealp{lrl,huub2}) and the radio centroid (e.g.
\citealp{bf.centroid,perryman.centroid}) have been found to be good
approximations of the core position of double radio sources.  We tested
their reliability, as well as the reliability for the point midway
between them, for 43 FR\,2 sources in the MS4 sample with
spectroscopically confirmed optical counterparts.

For each radio source we calculated the offsets between the optical
position and the following three positions: the MOST centroid, the
midpoint on the ATCA image, and the point midway between the MOST
centroid and the ATCA midpoint.  The MOST rather than ATCA centroid was
used because the ATCA images were often missing extended emission.
Median and mean distances from the candidate position to the search
position are listed in Table~\ref{tab4.radopt} for the three different
types of search position.  The mean and median distances to the
``midway point'' are the smallest, and it was therefore used as the
search position where possible.

In total 156 sources could be identified using the Schmidt plates.  Two
fields were obscured by stars; another two sources were unidentified
cluster sources (PKS~B1400$-$33 and MRC~B2006$-$566).  There were 54
blank fields. The remaining 14 sources had candidates selected from the
Schmidt plates which were later rejected on the basis of the CCD
images.

\subsection{Magnitudes from the COSMOS Database}

Optical $b_J$ magnitudes were obtained from Version~2 of the
UKST/COSMOS database of the southern sky
\citep{cosmos.yentis,cosmos.mjd}.  The bandpass $b_J$ for the UK
Schmidt southern sky survey lies roughly between $B$ and $V$ (e.g.
\citealt{iiiaj.evans}).  Errors in COSMOS magnitudes are typically
around 0.3 to 0.5\,mag \citep{cosmos.amu}.  We did not use COSMOS
magnitudes for 18 sources which were affected by blending.

The COSMOS classification of objects as stars or galaxies is often
inaccurate \citep{cosmos.amu}, sometimes leading to large photometric
errors.  The nonlinearity of plate opacity with source brightness,
and the differing light profiles of stars and galaxies, require COSMOS
to use different photometric algorithms for each \citep{edin.durham}.
We therefore applied the relevant algorithm to each candidate,
depending on our classification from visual inspection of the plates.
Photometric errors may still be large for objects with
uncertain morphology.

\subsection{Accuracy of Schmidt-plate Positions}
\label{sec.opt.acc}

For some objects, we used optical positions from digitised
Schmidt-plate data from COSMOS or the Digitized Sky Survey (DSS-I:
\citealp{access.dss,gsc.dss}).  We measured positions from the DSS-I
images in AIPS with a 9-pixel least-squares quadratic fit to the peak.

To estimate optical position accuracy, we compared DSS-I and BP
positions with VLBI radio positions \citep{vlbi.ma98} of
$\lesssim 10$\,milliarcsecond accuracy.  For DSS-I we used 14 MS4
quasars, 5 of which were on two or more DSS-I images.  The average
offsets in right ascension and declination were
$\overline{\Delta\alpha}=0''.01$ ($\sigma=0''.65$), and
$\overline{\Delta\delta}=0''.08$ ($\sigma=0''.64$) respectively.  For
the hand-measured BP positions we used 20 quasars from MS4 or the
Molonglo calibrator list \citep{rad.C1}.  The average offsets in right
ascension and declination were $\overline{\Delta\alpha}=0''.08$
($\sigma=0''.57$), and $\overline{\Delta\delta}=0''.32$
($\sigma=0''.58$) respectively.

COSMOS positions are affected by errors in the plate solutions
\citep{cosmos.mjd}, but nevertheless have good internal consistency
over small areas of sky, and being extracted from images with
relatively small pixel size, are more reliable than DSS for faint
objects.  We therefore used COSMOS positions to define the reference
frame in many of our CCD images (\S~\ref{aat.pos}).  Because
of the systematic errors, COSMOS positions of reference stars were
corrected wherever possible with DSS or BP positions.

To establish the systematic reliability of the optical positions in
the south, we compared BP and VLBI positions for 31 sources from the
MS4 sample or the Molonglo calibrator list.  A cross-plot is shown in
Figure~\ref{fig2.radopt}.  There was no observed systematic effect or
increased scatter in the errors of right ascension and declination as
a function of declination.

\subsection{Imaging at $R$ Band with the Anglo-Australian Telescope}
\label{aat.imaging}

Objects fainter than $b_J \sim 21$ were selected for imaging with
the 3.9\,m Anglo-Australian Telescope (AAT), to identify blank-field
sources, to classify faint candidates, and to obtain more information
about optical morphologies.  Observing was done at $R$-band to maximise
sensitivity to galaxies.  We used the $1024 \times 1024$ Tektronix CCD
detector at the f/3.3 prime focus to acquire images with a field size
of 6.7 $\times$ 6.7 arcmin$^2$ and a pixel size of 0\arcsec.39.
As the seeing was typically $1-2''$, the pixel images were well sampled.
The observing runs are summarised in Table~\ref{tab5.aatobs}.

Most of the observing was done in photometric conditions, except for
runs 3 and 4, which were affected by cloud or poor seeing for much of
the night.  Most objects from these runs were therefore re-observed in
run 6.  Two sources, MRC~B1358$-$493 and MRC~B1445$-$468, were observed
in a service run on 1994 February 14.  As well as the target sources,
dome and sky flat fields were observed.  Standard stars in the
E-regions \citep{e.reg} were observed at the beginning and end of each
night.

Exposures of 300 seconds were used; for a few faint objects we used
two or three exposures, and offset successive frames by 40\arcsec.
Guide stars were not used, but this generally did not lead to noticeable
tracking errors.

Bias subtraction and flat fielding were done using the STARLINK Figaro
software package.  After flat-fielding, multiple images of faint sources
were shifted and co-added using median offsets from about 10 stars in
the field.

Grey-scale plots of subsets of the CCD images are shown in
Figure~\ref{fig3.ccd}.  The images are centred at the optical
candidates, and have north to the top and east to the left.  Images of
some very faint candidates in the following fields have been smoothed
to improve the visibility of faint regions of emission:
MRC~B0008$-$421, MRC~B0315$-$685, MRC~B0615$-$365, MRC~B0647$-$475,
MRC~B1143$-$316, MRC~B1247$-$401, MRC~B1633$-$681, MRC~B1721$-$836,
MRC~B1756$-$663, and MRC~B1923$-$328.  We used {\sc ismooth} in Figaro
to perform a 9-point smooth, approximating the effect of convolution
with a Gaussian of FWHM 0\arcsec.5.

\subsection{Position Measurement}
\label{aat.pos}

Centroid positions of candidates were measured in FIGARO with {\sc cpos}
and {\sc centers}.  We typically used a circular aperture with radius of
3 to 5 pixels, depending on the seeing.  For a few extended candidates we
used a radius of up to 8 pixels.  We used COSMOS, BP, or DSS-I positions
of stars in the field to solve for equatorial positions as a function of
pixel position.  Because of systematic offsets in COSMOS, we then used the
BP or DSS-I positions of one or two brighter stars in the field to correct
the COSMOS-referenced positions.  The r.m.s value of the DSS-COSMOS
offsets was around 0\arcsec.9 in $\alpha$ and 1\arcsec.1 in $\delta$.

To check the accuracy of the CCD positions we compared optical
with VLBI positions for 6 sources: MRC~B0252$-$712, MRC~B0407$-$658,
MRC~B0615$-$365, MRC~B0647$-$475, \\
MRC~B1740$-$517, and MRC~B2259$-$375.
The VLBI-optical offsets are $\overline{\Delta\alpha}=0''.14$
($\sigma=0''.54$), $\overline{\Delta\delta}=0''.09$ ($\sigma=0''.58$).
The r.m.s offset is similar to the VLBI-BP and VLBI-DSS offsets found
in Section~\ref{sec.opt.acc}, indicating that the position error is
dominated by errors in the reference-star positions.

\subsection{Measurement of $R$ Magnitudes}
\label{sec.phot}

$R$ magnitudes were measured from the CCD images for the fainter galaxy
candidates.  It was not possible to do accurate photometry, because
standard stars had been observed only at the beginning and end of each
night, but the magnitudes obtained are still useful for providing rough
redshift estimates, as well as indicating the likely observing time
needed for spectroscopy.

No magnitudes were calculated for runs 3 and 4 which were affected by
cloud, nor for run~5, which had only two objects.  Photometric
solutions were kindly provided by Gary Da Costa for runs~1 and 2 and by
Tanya Hill for run~6.  The solutions are listed in Table~\ref{tab6.phot}.

Magnitudes were measured by summing pixels within a circular aperture
of radius 5~pixels ($\sim 2''$) using {\sc imexamine} in IRAF.  The fixed
aperture size may have led to an underestimation of the flux in cases
of poor seeing or very extended objects.  From repeat observations of
two fields, we estimate the error in our $R$ magnitudes to be about
0.5\,mag, about the same as the errors in the COSMOS magnitudes.  Most
of this is expected to be systematic error, with internal errors a
factor of 10 smaller.  For those candidates affected by blending with a
neighbouring star or galaxy, we made a rough magnitude estimate by
deblending the objects by eye.  These estimates have been flagged with
``:''.  The $R$ magnitudes are listed in column~6 of
Table~\ref{tab7.opt}.

\subsection{Likelihood of Chance Identifications}
\label{sec.likelihood}

We have estimated the number of spurious identifications among the 65
sources identified from the AAT images for which $R$ magnitudes have
been measured.  These are the objects for which such an estimate is
most useful, because of the substantial observing time required for
follow-up spectroscopy.  The estimated number of spurious
identifications, derived using the method of \citet{prob.downes},
came to about 13.

To make the estimate, source densities of galaxies in the field
were estimated from the $R$-band counts of \citet{br.cou2}.
Those of foreground stars were estimated from the star-count model of
\citet{bscou.80}, using the Fortran program \verb+model.f+ (J.\,Bahcall,
1997, {\it priv. comm.}), and assuming the following colour transformation
for giant and main-sequence stars:

\begin{equation}
R-V = - 1.576 (B-V) + 0.304,
\end{equation}

\noindent based on a least-squares fit to the colours measured by
\citet{starcol.83}.  Because of clustering, the errors in both star and
galaxy counts for any particular field are expected to be large.  Also,
the model of \citet{bscou.80} applies to fields with $|b| >
20^{\circ}$, whereas many MS4 sources have $10^{\circ} < |b| <
20^{\circ}$.  The estimate of 13 chance IDs is therefore uncertain.

\subsection{Estimation of Redshifts from Magnitudes}
\label{sec.estz}

Because few of the MS4 galaxies with $b_J > 19$ have measured
spectroscopic redshifts, it was necessary to use an alternative distance
estimate.  The most common method for radio galaxies is to use the fact
that their absolute magnitudes show a small scatter about a mean $M_V$
of $-$23 \citep{sandage.hd}.  The scatter in the relation between $K$
and $\log z$
is smaller than in the optical \citep{Rband.3cr}, but as $K$ magnitudes
were unavailable, $R$ and $b_J$ were used instead.  There will be large
uncertainties in the magnitude-based redshifts,  but they are still
useful for calculating parameters such as linear size and the $V/V_m$
ratio, which do not depend strongly on redshift.

We fitted a straight line to $\log z$ versus $R$ for 11 reliable
identifications, using an orthogonal least-squares fit \citep{isobe90}.
This gave a relation of

\begin{equation}
\log_{10}{z} = (-3.0 \pm 0.9) + (0.14 \pm 0.05)\,R
\label{eq.kpno}
\end{equation}

\noindent A plot of the fit is shown in Figure~\ref{fig4.magz}a.  As
there are few data points and large scatter, redshifts estimated from
this relation should be used with extreme caution.

As not all the galaxies had measured $R$ magnitudes, we fitted a line to
$\log_{10} z$ versus $b_J$ for the galaxies which were visible on the
Schmidt plates.  The fit was done using the ordinary-least-squares
bisector method \citep{isobe90}, and gave the result

\begin{equation}
\log_{10}{z} = (-3.95 \pm 0.18) + (0.17 \pm 0.01)\,b_J
\label{eq.bj}
\end{equation}

\noindent A plot of the fit is shown in Figure~\ref{fig4.magz}b.

Galaxy redshift estimates, based on equations~\ref{eq.kpno} or
\ref{eq.bj}, are given in parentheses in Table~\ref{tab7.opt}.  Their
errors are indeterminate, because of various factors which are hard to
quantify, such as dust extinction at high redshift.  The estimates are
provided purely as a guide, until full spectroscopic data can be obtained.
For quasar candidates we assumed $z=1$ when calculating linear sizes or
other physical quantities which do not strongly depend on redshift.

\subsection{Summary of Optical Data}
\label{sec.sumop}

Optical candidates have now been found for 97\% of the MS4 sample, on
the basis of the Schmidt plates or $R$-band CCD images.  Optical data
for each source are given in Table~\ref{tab7.opt}.

Candidates which appear compact on the Schmidt plates but have no
measured spectra have been flagged with ``Q?'' in Table~\ref{tab7.opt}.
Several candidates appear unresolved on the CCD but are faint or
invisible on the Schmidt plates.  If not foreground stars, these could
either be compact radio galaxies or reddened quasars.  These have been
given the flag ``Q?'' or ``g?'' in Table~\ref{tab7.opt}, depending on
the faintness of the optical candidate, and whether a radio core is
present.  Spectroscopy will be necessary to classify these objects with
more certainty.

Deeper CCD images are necessary for the two blank-field sources.
For around 20 sources with $R \gtrsim 22.5$, deeper CCD images would
be useful for giving more accurate optical positions.  IDs of 5 compact
radio sources are in doubt because of a significant radio-optical offset.
29 sources have extended radio structure and more than one optical
candidate; spectroscopy or higher-resolution radio images will be
necessary to confirm these identifications.

Apart from the two fields totally obscured by foreground stars, a
further 27 identifications are partially blended with stars.  Their
photometry and in a few cases their positional accuracy may be
affected.  Badly affected sources are discussed in more detail in the
comments in Section~\ref{sec.indcomm}.

The columns for Table~\ref{tab7.opt} are as follows:

\begin{enumerate}

\item IAU source name.  Other names are listed in the Comments on
Individual Sources.

\item Classification of the optical counterpart.

\item Right ascension of the optical counterpart, in J2000 co-ordinates.

\item Declination of the optical counterpart, in J2000 co-ordinates.

\item $b_J$ magnitude, usually from the COSMOS database.

\item $R$ magnitude, from the AAT CCD image.

\item Seeing of the AAT CCD image.

\item Redshift.  Values in parentheses are estimated from $R$ or
$b_J$, and are therefore uncertain.

\item References: for the first ID in the literature, finding chart, 
the optical position, the $b_J$ magnitude, and the redshift respectively.

\end{enumerate}

\section{COMMENTS ON INDIVIDUAL SOURCES}
\label{sec.indcomm}

In the comments below, radio-optical offsets use ATCA radio positions
from Table~\ref{tab3.atdat} in this paper and MOST radio positions from
Table~5 of Paper~I.  The abbreviation ``ACO89'' stands for the cluster
catalogue of \citet*{aco}.

\cind{MRC~B0003$-$567} Unusual S-shaped radio structure, suggestive of a
precessing jet.  The ATCA image contains strong sidelobes caused by poor
$uv$-coverage.  The tentative IDs of \citet{opt.w3} and \citet{opt.b2}
are rejected on positional grounds.

\cind{MRC~B0008$-$421} Gigahertz-Peaked-Spectrum (GPS) source.  A deeper
CCD image will be needed for this very faint ID.  The images of
\citet{opt.d8} and \citet{opt.d12} were blank to limiting magnitudes of
$V \sim 23$ and $i \sim 23$ respectively.  \citet{opt.v7} found
$R=24.3$, much fainter than our value of $R=22.6$.  One or both values
may be affected by error due to the faintness of the source, or the
source may be optically variable.

\cind{MRC~B0013$-$634} The optical field is partially obscured by the
$b_J=8.7$ star HD\,1208.

\cind{MRC~B0023$-$333} Wide-angle-tailed source; the VLA image
\citep{rad.E1} shows a radio core coincident with the optical
counterpart, the galaxy ESO\,350$-$G$-$15, brightest member of the
cluster AS\,41 (ACO89).

\cind{MRC~B0036$-$392} Quasar, blended with a foreground M star at
1\arcsec.5~E, 3\arcsec.3~S \citep{opt.t2}, and surrounded by several
faint objects, possibly an associated cluster.

\cind{MRC~B0042$-$357} Large ATCA-optical offset in right ascension
($1''.3$); a higher-resolution radio image is necessary to confirm the ID.

\cind{MRC~B0043$-$424} The redshift of 0.116 \citep{opt.t1} supersedes
that reported by \citet{opt.w6}.  The optical spectrum \citep{opt.t1}
shows weak [\ion{O}{2}] and [\ion{O}{3}] emission lines.  The object is
therefore a member of the class of radio galaxies with FR\,2 structure
and low-excitation optical spectra \citep{laing.94}.

\cind{MRC~B0048$-$447} The ID of \citet{rad.B2} is rejected on positional
grounds.

\cind{MRC~B0049$-$433} The fuzzy emission to the east and west of the
galaxy is roughly aligned with the radio axis.

\cind{MRC~B0103$-$453} Because the radio structure is extended and does not
show a core \citep{rad.J1}, a higher-resolution radio image will be
necessary for a secure ID.  Our candidate galaxy lies $17''$ from the
MOST centroid.

\cind{MRC~B0110$-$692} Blank field on the Schmidt plates.  The ID of
\citet{rad.J1}, a $b_J=21.7$ object, and the two other candidates
mentioned by them, all lie too far from the radio axis in a 5\,GHz ATCA
image (P.  Jones, 1992, {\it priv.  comm.\/}).  There is a possible
faint counterpart at $01^{\rm h}11^{\rm m}43\fs 08$
$-$68\arcdeg 59\arcmin 46\farcs 21 (J2000) on a
poor-seeing $R$-band CCD image.

\cind{PKS~B0131$-$36} S0 galaxy NGC\,612, studied in detail by
\citet{opt.e3}.  It contains a prominent dust lane, and is a rare
example of a radio galaxy with a disk.

\cind{MRC~B0214$-$480} ESO\,198$-$G$-$1, in the cluster AS\,239 (ACO89).

\cind{MRC~B0216$-$366} Stellar object, candidate (b) of \citet{opt.s7},
lies $5''$ from the MOST centroid.  A higher-resolution radio image is
necessary to confirm the ID.

\cind{MRC~B0223$-$712} The tentative ID of \citet{opt.l1} is rejected.

\cind{MRC~B0240$-$422} Asymmetric image; possible blend.  A $b_J=20.5$
galaxy, the candidate of \citet{opt.s5}, lying 24\arcsec.3~E,
22\arcsec.0~N of our ID, is closer to the radio centroid but further
from the midpoint.

\cind{MRC~B0251$-$675} The \citet{opt.s3} redshift ($z=2.11$) was based
on incorrect line identifications.  The ID of \citet{rad.P1} with a
cluster of galaxies is rejected on positional grounds.

\cind{MRC~B0252$-$712} The ID is blended with a foreground star at
2\arcsec.2~E, 1\arcsec.1~S (see \citealp{opt.t1}).

\cind{MRC~B0315$-$685} The large ATCA-optical offset of $1''.4$ may
be due to the extreme faintness of the optical counterpart; a deeper
optical image is needed.

\cind{PKS~B0319$-$45} The host galaxy ESO\,248$-$G$-$10 has a dust lane,
and coincides with a weak core in the ATCA image of \citet{rad.S15}.  A
detailed study of the host galaxy is given by \citet{opt.b22}.  The ID
of \citet{opt.t7} is rejected on positional grounds.

\cind{MRC~B0320$-$373} Well studied radio galaxy Fornax~A, identified with
the nearby elliptical galaxy NGC\,1316 \citep{opt.m4}.  The optical
position in Table~\ref{tab7.opt} is that of the nucleus \citep{opt.s28}.

\cind{PKS~B0332$-$39} ID first suggested by \citet{opt.b9}, and verified by
the radio images of \citet{rad.S14} and \citet{rad.E1}.  In the cluster
A\,3135 (ACO89).

\cind{MRC~B0336$-$355} Blended at low resolution with the weaker
radio source MRC~B0336$-$356, and at first wrongly identified with
that source's optical counterpart, the galaxy NGC\,1399 \citep{rad.M1}.

\cind{MRC~B0344$-$345} In a sparse cluster behind the Fornax cluster.
Several other galaxies lie within the bounds of the radio structure,
including a $b_J=17.0$ elliptical galaxy coincident with the peak in the
eastern lobe \citep{rad.S14,rad.E1}.

\cind{MRC~B0357$-$371} A $b_J=19.5$ galaxy at 1\arcsec.7~W,
17\arcsec.6~S was suggested as the ID by \citet{rad.B2}, \citet{opt.w1},
and \citet{rad.J1}, but lies outside the radio structure in an ATCA image
(R. Subrahmanyan, 1992, {\it priv.  comm.\/}), and is therefore rejected.

\cind{MRC~B0407$-$658} The tentative IDs of \citet{opt.h4} and \citet{opt.p8}
are rejected on positional grounds.

\cind{MRC~B0409$-$752} Our ATCA image has poor hour-angle coverage; the
8.6\,GHz image of \citet{map.0409} has higher resolution and dynamic
range.  The tentative ID of \citet{opt.w3} is rejected on positional
grounds, and the IRAS source B04099$-$7514 is associated not with the
radio source but with a galaxy in a foreground cluster.

\cind{MRC~B0411$-$647} `Fat double' with extended lobes; a faint possible
core lies close to the galaxy ID.  Probably associated with the cluster
A\,3231 (ACO89).

\cind{MRC~B0420$-$625} The ID of \citet{opt.l1} and \citet{opt.s4} is
rejected on positional grounds.

\cind{MRC~B0427$-$366} The central peak of the radio triple is likely to be
a hot spot rather than the core, as no object coincides with it on the
CCD image.  The probable ID, a stellar object, lies further to the
south.  \citet{opt.s16} note this candidate but mark the wrong object on
their finding chart.

\cind{PKS~B0427$-$53} Dumb-bell galaxy IC\,2082, the brightest member of
the small cluster AS\,463 (ACO89).  Identified by Shapley \citep{opt.b20}.
\citet{map.interaction} find that the radio source is associated with
the fainter optical nucleus.

\cind{MRC~B0429$-$616} The ID of \citet{rad.J1} is rejected on the basis of
an ATCA image \citep{areid.thesis}, showing a WAT structure with a
radio core coinciding with the ID of \citet{opt.u1}.  In the cluster
A\,3266 (ACO89).

\cind{MRC~B0436$-$650} The radio image has low dynamic range.  The candidate
ID is a faint diffuse galaxy which lies on the radio axis and close to
the radio centroid.  The stellar ID of \citet{opt.s4} is rejected.

\cind{MRC~B0456$-$301} A VLA image (A. Unewisse, 1993, {\it priv. comm\/.})
reveals an amorphous structure with no core or hot spots; possibly a fat
double viewed end-on.  In the cluster A\,3297 (ACO89).

\cind{MRC~B0509$-$573} Stellar object, confirmed as a quasar in an
unpublished AAT spectrum obtained by R.W.H.

\cind{PKS~B0511$-$30} Galaxy, AM\,0511$-$303, coincident with the core in
the image of \citet{rad.E1}.  The optical morphology is very disturbed
and peculiar \citep{rad.S16}.  Candidates $b$ and $c$ of \citet{opt.s7}
are rejected.

\cind{MRC~B0511$-$484} The long wisps described by \citet{opt.r1}, and
interpreted as tidal debris, appear clearly in the CCD image
(Figure~\ref{fig3.ccd}).

\cind{MRC~B0513$-$488} Stellar object, confirmed as a quasar in an
unpublished AAT spectrum obtained by R.W.H.

\cind{MRC~B0521$-$365} Well studied radio source \citep{rad.E1,map.keel},
identified with ESO\,362$-$G$-$21.  The ID has been classified at
different times as an N~galaxy \citep{opt.b9} and as a BL~Lac object
\citep{bllac.araa}, as both the continuum \citep{eggen70,shen72} and
spectral lines \citep{ulrich81} are highly variable.  The redshift of
0.061 found by \citet{opt.w5} is incorrect, possibly because the
emission lines were diluted at the time of measurement \citep{opt.s9}.

\cind{MRC~B0602$-$647} Dumb-bell galaxy; the radio centroid is slightly
closer to the south-eastern member of the pair.

\cind{MRC~B0618$-$371} Dumb-bell galaxy, ESO\,365$-$IG$-$6.
In the VLA image of \citet{rad.E1} the radio core
is between the two galaxies, but in that of \citet{map.parma} it
coincides with the eastern galaxy.

\cind{MRC~B0620$-$526} The optical field is partially obscured by
Canopus.

\cind{MRC~B0625$-$354} Dumb-bell galaxy, in the cluster A\,3392 (ACO89).
Partially confused on the Schmidt plates by two foreground stars
\citep{opt.p8}.

\cind{MRC~B0625$-$536} Dumb-bell galaxy ESO\,161$-$IG$-$7, the brightest
member of the cluster A\,3391 (ACO89).  An ATCA image \citep{map.dumbbell}
shows that the radio core coincides with the eastern galaxy of the pair.

\cind{MRC~B0625$-$545} In the cluster A\,3395 (ACO89).

\cind{MRC~B0658$-$656} The optical field is obscured by a $b_J=14.5$ stellar
object, suggested as the ID by \citet{opt.w15}, but found to be a
foreground star by \citet{opt.t2}.

\cind{MRC~B0704$-$427} Stellar object, confirmed as a quasar in an
unpublished AAT spectrum obtained by R.W.H.

\cind{PKS~B0707$-$35} Galaxy, with an extended asymmetric halo, coincides
with the radio core.  The ID of \citet{opt.s7} is rejected on positional
grounds.

\cind{MRC~B0743$-$673} The redshift of 0.395 \citep{opt.t5} is incorrect
\citep{opt.j6}.

\cind{MRC~B0842$-$754} Unusual source with radio lobes roughly perpendicular
to one another.  The radio peak in the ATCA image lies near the optical
counterpart, which shows an apparent extension to the north-east on the
CCD image (Figure~\ref{fig3.ccd}).  \citet{opt.t1} comment that the
emission lines may be extended.

\cind{MRC~B0846$-$811} The ID consists of two faint emission regions,
aligned with the radio axis: they could be unrelated galaxies, or part
of the same galaxy.  The position given in Table~\ref{tab7.opt} is that
of their centroid.

\cind{MRC~B0906$-$682} The optical field is partially obscured by the star
HD\,78913.

\cind{MRC~B0943$-$761} The galaxy ID appears to be the cD of a cluster.
This is supported by its detection as a ROSAT source \citep{rosat.id}.

\cind{MRC~B1030$-$340} Partially blended with another galaxy 1\arcsec.4\,E
which shows wispy trails, suggesting a tidal interaction.  The galaxy
pair appears on the Schmidt plate as a single $b_J=20.7$ galaxy, the ID
of \citet{opt.d14}.

\cind{MRC~B1036$-$697} The candidate described by \citet{opt.h7} as
`possibly an unusual galaxy' appears as two stellar objects on the
Schmidt plate; the correct ID is to the south of these.

\cind{MRC~B1056$-$360} The elliptical galaxy ID coincides with a weak radio
core in the image of \citet{rad.E1}.

\cind{MRC~B1123$-$351} Elliptical galaxy ESO\,377$-$G$-$46, in the cluster
AS\,665 (ACO89).

\cind{MRC~B1136$-$320} Our proposed ID lies on the radio axis and coincides
with a possible core in the image of \citet{map.sproats}.  The ID of
\citet{rad.J1} is rejected.

\cind{MRC~B1151$-$348} Quasar, $z=0.258$, appears slightly non-stellar
on the Schmidt plates.

\cind{MRC~B1221$-$423} Unusual galaxy, with a bright nucleus, diffuse
envelope, and a blue `knot' of more compact emission to the south
\citep{opt.s38,opt.j10}.

\cind{MRC~B1232$-$416} The large ATCA-optical offset of 1\arcsec.8
may mean this ID is incorrect.  The IDs suggested by
\citet{opt.s1} and \citet{opt.l1} are rejected on positional grounds.

\cind{MRC~B1234$-$504} The ID is uncertain: the large radio-optical
offset of 2\arcsec.5, together with the crowded field at low Galactic
latitude, could indicate that the candidate is a foreground star.
Deeper observations in good seeing are needed.

\cind{MRC~B1243$-$412} Stellar object, confirmed as a quasar in an
AAT spectrum obtained by the authors.

\cind{MRC~B1246$-$410} Compact steep-spectrum source with unusual
Z-shaped radio structure, possibly a short wide-angle-tail source
\citep{burns.wat}.  The largest angular extent of
32\arcsec\ corresponds to 6\,kpc, well within the host galaxy, the well
studied elliptical galaxy NGC~4696, and brightest member of the
Centaurus cluster A\,3526 (ACO89).

\cind{MRC~B1259$-$769} The ID is unusually faint.  The fuzzy object at
4\arcsec.3~W, 0\arcsec.9~N showed no strong emission lines in an AAT
spectrum, and is much redder.  The two objects are aligned with the radio
structure; the redder object may be a companion galaxy, or possibly a
foreground object.

\cind{MRC~B1259$-$445} Several stellar objects and faint galaxies lie near
the radio position.  The most likely candidate is the brightest galaxy
in a small group.  The ID of \citet{rad.J1} is probably a foreground star.

\cind{MRC~B1302$-$491} Edge-on spiral NGC\,4945.  The optical properties
have been studied in detail by \citet{opt.d13}, \citet{opt.p12},
and others.

\cind{MRC~B1303$-$827} Stellar object, confirmed as a quasar in an
unpublished AAT spectrum.

\cind{PKS~B1318$-$434} Elliptical galaxy NGC~5090, interacting with the
spiral galaxy NGC~5091; studied in detail by \citet{opt.s30}.

\cind{MRC~B1322$-$427} Well studied low-luminosity radio source
Centaurus~A, the nearest radio galaxy, reviewed by \citet{cena.review}
and \citet{cena.review2}.  The optical counterpart is the dust-lane
elliptical galaxy NGC~5128.

\cind{MRC~B1330$-$328} The radio structure appears similar to a head-tail
source, but the confirmed optical counterpart, a galaxy, lies between
the mid-point and the centroid, rather than at one end of the structure.
The source is probably a highly asymmetric FR\,2 double.  The ID of
\citet{opt.g2} is too far south of the radio position, and that of
\citet{opt.p3} was found to be a foreground star by \citet{opt.j5}.  The
correct ID is just to the east of this star, not to the west as stated
by \citet{opt.s15}.

\cind{PKS~B1333$-$33} Galaxy IC\,4296, brightest member of the cluster
A\,3565 (ACO89).

\cind{MRC~B1346$-$391} Very faint diffuse ID is partly blended with a
stellar object at 2\arcsec.2~W, 1\arcsec.4~S, and a brighter stellar
object at 3\arcsec.7~W, 0\arcsec.2~S.

\cind{PKS~B1400$-$33} Probable relic radio source associated with the
poor cluster around NGC\,5419.  Refer to \citet{rad.S17} for the most
up-to-date interpretation of this enigmatic source.

\cind{MRC~B1407$-$425} Unusual amorphous radio source, consisting of
extended emission confined within 50\arcsec\ (about 50 kpc).  Identified
with elliptical galaxy ESO\,271$-$G$-$20.  The ATCA image contains a
deep negative bowl, because of the sparse coverage at short baselines.

\cind{MRC~B1445$-$468} ID is partially obscured by a $b_J=16.6$ stellar
object at 0\arcsec.5~W, 2\arcsec.4~N.  The galaxy position could not be
measured accurately, but appears to coincide with a probable core in
the ATCA image.

\cind{MRC~B1451$-$364} The ID of \citet{rad.J1} at 18\arcsec.3~E,
13\arcsec.8~S appears stellar on the CCD, and is too far from the radio
axis.  Our ID is one of their other candidates.

\cind{MRC~B1526$-$423} Galaxy, partially blended with a star
2\arcsec.1~W.

\cind{MRC~B1540$-$337} Galaxy, partially obscured by a $b_J=16.5$ stellar
object at 3\arcsec.4~E, 1\arcsec.2~N,  suggested as the ID by \citet{opt.j1},
but found later to be a foreground star \citep{opt.j2,opt.t2}.

\cind{MRC~B1540$-$730} Compact object, 6\arcsec.7 from the MOST centroid.

\cind{MRC~B1549$-$790} Galaxy, candidate (b) of \citet{opt.p8}, is
extended north-south on the CCD image, and may be interacting with a close
companion.  Detected in the IRAS Point Source Catalog \citep{iras.psc}.

\cind{MRC~B1610$-$771} Quasar with a flat radio spectrum and steep optical
spectrum \citep{opt.h6,opt.c7}, blended by COSMOS with a foreground star
\citep{opt.c7} at 3\arcsec.0~W, 3\arcsec.5~N.  $b_J$ was estimated from
SuperCOSMOS magnitudes of stars of similar brightness in the same field
\citep{supercos}.

\cind{MRC~B1622$-$310} The large ATCA-optical offset of 3\arcsec.7 is
probably due to poor phase calibration of the ATCA image, as there is
good agreement between the MOST and optical positions.

\cind{MRC~B1633$-$681} Extended radio source in a crowded optical field -- a
higher-resolution radio image will be necessary for an unambiguous ID.
Tentative ID with faint galaxy, 9\arcsec.7 from the radio centroid.

\cind{MRC~B1637$-$771} The redshift of 0.024 reported by \citet{opt.d2}
may be a typographical error, because the redshifts of \citet{opt.b17},
\citet{opt.t1} and \citet{opt.s15} agree with the value of 0.0423 found
by \citet{opt.w6}.

\cind{MRC~B1655$-$776} The redshift of 0.0663 \citep{opt.w6} is
inconsistent with the later value of $z=0.0944$ \citep{opt.s15}, which
we assume to be correct.

\cind{MRC~B1716$-$800} The ID of \citet{rad.P1} with the spiral galaxy
IC\,4640 is rejected on positional grounds.

\cind{MRC~B1721$-$836} The optical candidate is extremely faint: a deeper
optical and a higher-resolution radio image are necessary for a secure
ID.

\cind{MRC~B1740$-$517} The two candidates of \citet{opt.p8} are rejected on
positional grounds.  The images on the CCD are slightly elongated
east-west, owing to tracking errors, and are affected by
scattered light from a bright star to the south-west.

\cind{MRC~B1754$-$597} The optical object has a faint wisp of emission
extending about 4\arcsec\ to the south-west along a position angle of
$-108^{\circ}$.  The candidates of \citet{opt.b2} and \citet{opt.h7}
are rejected on positional grounds.

\cind{MRC~B1756$-$663} Faint galaxy.  The ID of \citet{opt.s1} is
rejected on positional grounds.

\cind{MRC~B1814$-$519} The candidate of \citet{opt.h7}, a $b_J=15.8$
stellar object, was found to be a foreground star by \citet{opt.t2}.

\cind{MRC~B1814$-$637} Galaxy, partially obscured by a foreground star;
ID wrongly rejected by \citet{opt.t6}.

\cind{MRC~B1817$-$391} Very crowded optical field: ID is uncertain.

\cind{MRC~B1818$-$557} Marked galaxy is the most plausible ID for this
extended radio source.

\cind{MRC~B1840$-$404} As there is no visible radio core
\citep{map.sproats}, and several candidates lie within the radio
structure, the ID remains tentative.

\cind{MRC~B1853$-$303} Proposed ID is compact and lies 1\arcsec.7 from the
radio centroid.

\cind{MRC~B1917$-$546} The ATCA snapshot image has low dynamic range
because of the weakness of the source.  There is very faint diffuse
emission at around $19^{\rm h}21^{\rm m}52\fs 88$,
$-$54\arcdeg 31\arcmin 54\farcs 0 (J2000), near the radio core in an
unpublished 1.4\,GHz ATCA image.  A deep image in good seeing is required.

\cind{MRC~B1922$-$627} The ID of \citet{rad.J1} was found to be a
foreground star in an AAT spectrum obtained by the authors.

\cind{MRC~B1923$-$328} A deeper optical image is necessary for a secure ID.

\cind{MRC~B1929$-$397} Dumb-bell galaxy ESO\,338$-$IG$-$11, brightest
member of the poor cluster AS\,820 (ACO89).  The radio core in the
image of \citet{rad.E1} coincides with the brighter, south-east member
of the pair.

\cind{MRC~B1932$-$464} \citet{opt.p8} wrongly rejected this ID.

\cind{MRC~B1933$-$587} An AAT spectrum taken on 1989 August 03 shows narrow
lines of Ly$\alpha$ and \ion{C}{4}, characteristic of other CSS quasars
observed by \citet{composite.jcb}.

\cind{MRC~B1940$-$406} In the cluster A\,3646 (ACO89).  The two candidates of
\citet{opt.s7} are rejected on positional grounds.

\cind{MRC~B1953$-$425} The nearby $b_J=16.8$ stellar object suggested
as the ID by \citet{opt.h5} was found to be a foreground star by
\citet{opt.j5}.

\cind{MRC~B1954$-$552} Bright elliptical galaxy, partially obscured by
a bright foreground G0 star \citep{opt.w3}.

\cind{MRC~B2006$-$566} Very extended, diffuse ultra-steep-spectrum radio
source, associated with the cluster A\,3667 (ACO89).  This source belongs
to the rare class of diffuse relic sources associated with clusters but
not with any particular galaxy.  The ID of \citet{rad.J1} is rejected
on the basis of the ATCA image of \citet{rad.R2}.

\cind{MRC~B2009$-$524} The ATCA image has very low dynamic range.

\cind{MRC~B2020$-$575} Compact galaxy, possibly blended with a fainter
object to the south.

\cind{MRC~B2028$-$732} The ID of \citet{rad.P1} with IC\,5016 is rejected
on positional grounds.

\cind{MRC~B2031$-$359} Dumb-bell galaxy ESO\,400$-$IG$-$40, the dominant
member of the cluster A\,3695 (ACO89).  The radio source is identified
with the northern, slightly fainter member of the pair \citep{rad.B2}.

\cind{MRC~B2032$-$350} Galaxy, candidate 2 of \citet{opt.d8}.  The two faint
candidates of \citet{opt.p8} are rejected on positional grounds.

\cind{MRC~B2041$-$604} The tentative ID of \citet{opt.w3} is rejected on
positional grounds.

\cind{MRC~B2049$-$368} The ID of \citet{opt.s4} is rejected on positional
grounds.

\cind{MRC~B2052$-$474} The tentative ID of \citet{opt.t7} is rejected on
positional grounds.

\cind{MRC~B2122$-$555} Unusual radio triple.  The highly asymmetric
structure is similar to that of the $z=0.371$ quasar 3C~351 (see
\citealt{qpol.3freq}).

\cind{MRC~B2140$-$434} The extended radio emission is poorly imaged
but the ID is secure.

\cind{PKS~B2148$-$555} In the cluster A\,3816 (ACO89).

\cind{MRC~B2153$-$699} Galaxy, object A of \citet{map.scm}, identified by
\citet{rad.J1}.

\cind{MRC~B2201$-$555} The tentative ID of \citet{opt.b2} is rejected on
positional grounds.

\cind{MRC~B2213$-$456} The IDs of \citet{opt.l1}, \citet{opt.s4}, and
\citet{opt.b2} are rejected on positional grounds.

\cind{MRC~B2223$-$528} Possible tidal interaction with neighbouring galaxy
to south east.

\cind{MRC~B2226$-$386} The $b_J=16.4$ stellar object at 7\arcsec~W is a
foreground star \citep{opt.t5}.

\cind{MRC~B2250$-$412} Higher-resolution radio image in
\citet{map.morganti}.  There is an extended emission-line region
adjacent to the western lobe \citep{2250.arc}.

\cind{MRC~B2319$-$550} The ID of \citet{opt.b7} is rejected on positional
grounds.

\cind{MRC~B2323$-$407} Faint clumpy galaxy, roughly aligned with the radio
axis.

\cind{MRC~B2332$-$668} A $b_J=12.6$ foreground star obscures the optical
field.  The tentative ID of \citet{opt.b2}, a spiral galaxy, is rejected
on positional grounds.

\cind{MRC~B2354$-$350} Galaxy, ESO\,349$-$G$-$10, brightest member of
the cluster A\,4059 (ACO89), which contains an extended X-ray source,
peaked on the radio source (see \citealt{heinz.abell4059}).

\section{SUMMARY OF SAMPLE CONTENT}
\label{sec.sampcont}

We now summarise the global properties of the MS4 sample, and compare them
with those of the 3CRR, the northern equivalent.  For this purpose we
also used the SMS4 sample, a subset of MS4 defined in Paper~I to have
the same flux-density cutoff, $S_{\rm 178\,MHz} = 10.9$\,Jy, so that we
could test the uniqueness and completeness of the 3CRR sample.  In the
future, when more complete angular size and redshift information is
available for SMS4, it may be advantageous to combine the 3CRR and SMS4
samples, to improve the statistical accuracy of the derived parameters.

Source numbers and median angular size, redshift ($z$), linear size
($l$), and radio power ($P_{408}$) for the MS4, SMS4, and 3CRR samples
are summarised in Table~\ref{tab8.medians}.  We used cosmological
parameters of $H_0=71\,{\rm km\,s^{-1}\,Mpc^{-1}}$, $\Omega_m=0.3$,
$\Omega_{\Lambda}=0.7$ \citep{wmap.param} for these and subsequent
calculations, unless otherwise stated.  Errors of the median were
estimated using the method described by \citet{medianerr.cj}.  Because
fewer than half the MS4 sources have spectroscopic data, the median
redshift, linear size, and radio power are uncertain, but even so the
median radio power is very similar for SMS4 and 3CRR.  The median
redshifts for 3CRR and SMS4 are very similar, with the MS4 sample
having a slightly higher median redshift, as expected from its lower
flux-density limit.  The slightly higher median linear size of SMS4
than of 3CRR may not be significant, because of the large errors.

The identification content of the three samples is summarised in
Table~\ref{tab9.ngq}.  Taking into account unconfirmed candidates, the
fraction of quasars in MS4 is $(18 \pm 3)$\%, compared with $(25 \pm
3)$\% for 3CRR.  The slightly lower proportion of quasars in MS4 is
unexpected, given that the quasar fraction of flux-density-limited
samples tends to increase with frequency, but the difference is barely
significant.

Table~\ref{tab10.struc} contains a brief summary of radio-structure
information obtained from other observations besides those in this paper.
Column~2 gives the Fanaroff-Riley (or other) structural classification:
codes are explained in the footnote to the table.  Columns~3 and 4 give
the largest angular extent and position angle of extension (in degrees
east of north) respectively, and column~5 the reference to the radio image
from which these values were measured.  The largest angular extent was
measured by preference from high-resolution images for compact sources
or sources with edge-brightened structure, and from lower-resolution
images for sources with extended edge-darkened structures.

The Fanaroff-Riley structural classifications are summarised in
Table~\ref{tab11.fr}.  The radio structures were classified on the basis
of the ATCA 5\,GHz images or other radio images from the literature
(references in Table~\ref{tab10.struc}).  Fat doubles have been classed
with FR\,1 sources \citep{ol89}, as have sources of borderline
FR\,1/FR\,2 category.  D2 sources and triple sources with
edge-brightened lobes have been classified as FR\,2.  Of the
unclassified MS4 sources, most do not have images which resolve their
structure; five are well resolved but show structure which cannot be
readily classified as FR\,1 or 2.

A detailed comparison of radio structures cannot be made until
higher-resolution radio images are available for the unclassified MS4
sources, i.e. those which lie within specific ranges of angular size
(LAS $\lesssim 5''$, 35\arcsec\ $<$ LAS $<$ 75\arcsec).  The median
angular size of the unclassified sources is around 2\arcsec, much less
than the median for the whole MS4 sample of 27\arcsec.  Given that
FR\,2 sources in flux-density-limited samples have on average smaller
angular size, an unclassified source is therefore much more likely to
have FR\,2 structure.  The smaller angular extent can be explained by
the higher radio luminosity of FR\,2 sources, which because of
Malmquist bias will ensure that FR\,2 sources lie on average at higher
redshift than sources with FR\,1 structure.  This expectation is
supported by the fact that the proportions of sources classified as
FR\,1 are similar in the SMS4 and 3CRR samples, suggesting that most of
the SMS4 FR\,1 sources are likely to have been classified as such
already.

A colour-colour plot ($\alpha_{843}^{2700}$ versus
$\alpha_{408}^{843}$, defined in the sense $S_{\nu} \propto
\nu^{\alpha}$) for the MS4 sample is shown in Figure~\ref{fig5.col}.
More sources tend to lie below the diagonal than above, indicating that
on average spectra tend to steepen at high frequency.  The median
values of $\alpha_{408}^{843}$ and $\alpha_{843}^{2700}$ are $-0.83 \pm
0.02$ and $-0.91 \pm 0.01$ respectively.

The MS4 sample contains 41 CSS sources, here defined to have linear
size $<25$\,kpc, and spectral index $\alpha_{408}^{2700} < -0.5$.
Those CSS not listed in Table~\ref{tab3.atdat} are the ATCA calibrators \\
MRC~B1234$-$504 and MRC~B2259$-$375.  The two GPS sources in the MS4
sample are \\
MRC~B0008$-$421 and MRC~B1934$-$638.  The SMS4 sample
contains no GPS sources; their absence is not surprising as these
sources are intrinsically rare \citep{odea98.gps}, and with spectra
that turn over at $\sim$1\,GHz, are likely to be weak at 178\,MHz.

\subsection{Apparent Magnitudes and Quasar Redshifts}
\label{sec.appmag.dist}

Histograms of the $b_J$ magnitudes for galaxies and quasars in the MS4
sample are shown in Figure~\ref{fig6.histbj}.  The distribution for quasars
peaks at around $b_J=18.5$, well above the plate limit.  In contrast,
the galaxy number counts increase up to the plate limit.  Comparison
with the 3CRR is not possible at present, as a complete set of $b_J$,
$B$ or $V$ magnitudes is not yet available for the entire 3CRR sample.

We have, however, compared the magnitudes of MS4 quasars with those of
another radio-selected quasar sample.  The median $b_J$ for MS4 quasars
is \merr{17.6}{-0.5}{+0.4}, compared with $18.8 \pm 0.2$ for the
Molonglo Quasar Sample (MQS: \citealp{mqs.id,mqs.spectra}), which was
selected to have $S_{408} > 0.95$\,Jy.  The difference in median
magnitudes is consistent with a weak correlation between radio
and optical luminosities which has been found for steep-spectrum quasars
\citep{q.radoptcor}.  The cause of the correlation is not clear, but it
is not dominated by effects of redshift or jet orientation
\citep{q.radoptcor}.

The median redshifts of quasars in the MS4, SMS4, and 3CRR samples are
listed in Table~\ref{tab12.medz}.  A Kolmogorov-Smirnov two-sample test
of the redshift distributions for SMS4 and 3CRR quasars does not show
a significant difference at the 10\% level.  A similar null result was
obtained for a comparison between MS4 and 3CRR.

\subsection{The $V/V_m$ Test}
\label{sec.vvm}

The $V/V_m$ test \citep{vvm,vvm.rr} is a method of studying the space
distribution of a flux-density-limited sample.  For each source, the
volumes $V$ and $V_m$ of two spheres centred at the observer are
calculated: the first sphere with the source on its surface, and the
second extending to the distance at which the source's flux density
would be equal to the sample's limit.  The ratio $V/V_m$ always lies
between 0 and 1.  For a sample uniformly distributed in space and with
no luminosity evolution the mean value of $V/V_m$ should be near 0.5.
For samples of steep-spectrum quasars, such as those in 3CR, the mean
value of $V/V_m$ is around $0.67 \pm 0.05$ \citep{mw.vvm,schmidt.vvm},
suggesting an excess at high redshift.  For samples of flat-spectrum
quasars the mean $V/V_m$ is smaller, around 0.57 \citep{schmidt.vvm},
suggesting less cosmological evolution.  \citet{lrl} compared the
distributions of $V/V_m$ for the radio galaxies and quasars in the 3CRR
sample, and found no significant differences between them at the 20\%
level.  This is consistent with radio galaxies and radio quasars being
different forms of the same kind of object.

In order to do a similar test, we calculated $V/V_m$ for the radio
galaxies and quasars in the MS4 sample.  Because the samples were
almost completely optically identified, only the radio flux-density
limit of 4.0\,Jy was considered, and no optical cutoff was used.  For
the 117 MS4 sources without spectroscopic redshifts, we used redshifts
estimated from $R$ or $b_J$ magnitudes (\S~\ref{sec.estz}).  The
value of $V/V_m$ is not a strong function of estimated redshift.
Table~\ref{tab13.vvm} shows that $V/V_m$ is very similar for galaxies in
each of the three samples.  For quasars the SMS4 has a slightly larger
mean $V/V_m$ than the 3CRR, but the difference is not significant at
the $2\sigma$ level.

\subsection{Tests of Unified Models}

The low selection frequencies of the MS4 and SMS4 samples make them ideal
for testing unified schemes of powerful radio sources.  In these models
(see reviews by \citealp{unif.antonucci}, \citealp{unif.urrypad},
\citealp{case.unif}, and \citealp{review.wills99}), the observed
properties of radio-loud active galaxies are assumed to be a strong
function of the orientation of the radio jet with respect to the
observer.  At small viewing angles, core and jet flux densities will be
enhanced by Doppler boosting \citep{doppboost.bk}.  The only viable
candidates for the corresponding unbeamed population at larger viewing
angles are the powerful radio galaxies
\citep{rgpar.owen86,rgpar.scheu87,rgpar.pea87}, with all radio quasars
probably beamed versions of the same objects \citep{barthel.qbeam}.
The difference in their optical properties could be explained by a
dusty torus obscuring the broad-line regions and quasar continuum
\citep{barthel.qbeam}.

Provided that the small- and large-scale radio structures are aligned,
projected linear sizes of the unbeamed population should be larger on
average than those of the beamed objects.  As the steep-spectrum
emission from the radio lobes is unlikely to be affected by beaming,
samples selected at low radio frequency should be relatively free of
orientation bias and therefore ideal for testing unified models.
\citet{barthel.qbeam} tested his model on a subset of the 3CRR sample
with $0.5 < z < 1$, comparing the numbers and projected linear sizes of
the 12 quasars and 30 powerful ($P_{178} \gtrsim
10^{28}$\,W\,Hz$^{\rm -1}$) radio galaxies in this redshift range.
Barthel excluded the CSS sources from the comparison, but including
them did not have a large effect on the results.  From the number ratio
he inferred that sources viewed at angles of $\psi < 44^{\circ}.4$ to
their jet direction were quasars, and those at larger $\psi$ were
galaxies.  This value was consistent with the ratio of median linear
sizes of the quasars to that of the radio galaxies.

However, the study of \citet{singal93a} of 3CRR sources at other
redshift ranges, and sources from the 1\,Jy sample of \citet{1jy.1},
cast doubt on this unified model, because at other redshift ranges, the
ratios of quasar to galaxy linear sizes were different.  Furthermore, a
study of a 408\,MHz-selected 1\,Jy sample of around 550 sources
\citep{iau95.linsiz} found that for $z > 0.5$, the median linear size
for quasars is actually larger than for galaxies.  Because of the
difficulty of interpreting these results in terms of a simple
orientation picture, it was important to perform similar tests on
another sample.  Given the comparative rarity of powerful radio
sources at any redshift range, a southern sample such as the MS4 is
valuable in supplementing the 3CRR data, despite the fact that the MS4
sample does not yet have a uniform set of spectroscopic information,
core flux densities, or high-resolution radio images.  Although some of
our redshifts are magnitude-based estimates (\S~\ref{sec.estz}),
in standard cosmological models the ratio of linear to angular size
does not vary greatly over the expected range of redshift.

Of the galaxies with high-resolution radio images, only those with
unambiguous FR\,2 structure have been considered.  Extended sources
without high-resolution radio images have been classed as FR\,2 unless
there is evidence to the contrary.  To avoid contamination from
core-dominated sources, we considered only those sources with 408\,MHz
{\it extended} emission above the flux-density limit.  We estimated the
extended value of $S_{408}$ by assuming the core had a flat radio
spectrum ($\alpha=0$), and subtracting the core flux density at 2300 or
5000\,MHz from the total 408\,MHz flux density.  CSS sources were also
excluded for consistency with previous work: it is not certain how the
CSS sources may fit in to unified schemes.

In Table~\ref{tab14.qf}, linear sizes of MS4 galaxies and quasars
in the range $0.5 < z < 1$ are compared with the 3CRR sample.  The
numbers are slightly different from those of \citet{barthel.qbeam}
because of revisions to 3CRR and because of the different cosmology
used here.  We have excluded from the comparison galaxies known to have
low-excitation optical spectra, as they may be lacking a quasar nucleus
\citep{laing.94}.  The quasar fractions of MS4 and SMS4 are similar to
those of 3CRR, but the ratios of quasar to galaxy linear sizes are
larger in the southern samples.  This may be due to the smaller number
of quasars, but it certainly leads one to question the reliability of
the earlier interpretation.

To increase the sample size, median linear sizes for the entire redshift
range were then compared with two other samples for quasars and one other
for galaxies (Table~\ref{tab15.lin}).  Aside from the SMS4 sample, with
only 14 quasars and therefore large uncertainty, the median quasar sizes
are larger in the samples selected at lower flux density.  On the other
hand, the median galaxy sizes do not vary significantly with sample flux
density limit.

The model of \citet{linsiz.gkkw} accounts for the discrepant linear-size
ratios by invoking not only orientation, but temporal evolution of
radio linear size and luminosity.  Their model also assumes that the
critical viewing angle $\psi$, within which the source's observed
optical properties are those of a quasar, is larger for sources with
greater initial radio power $P_0$.  Samples such as 3CRR, containing
many sources of high radio luminosity, would be dominated by young
sources of high $P_0$.  Given the similar ages and initial luminosities,
orientation effects would dominate, with galaxies being larger than
quasars.  Samples of lower flux density would contain a greater mix
of ages and $P_0$ values, with old sources of high $P_0$ likely to be
quasars of large linear size, and young sources of low $P_0$ likely to
be galaxies of small linear size.  This could explain the comparatively
large quasar sizes in the 1\,Jy sample of \citet{iau95.linsiz}.
\citet{linsiz.gkkw} found the relation between linear-size ratios and
quasar fractions for several low-frequency samples to be consistent
with their predictions.  The MS4 is a useful addition to their data as
it occupies a flux-density range between that of the 3CRR and those of
weaker samples.  For the 131 relevant sources from the MS4, the ratio of
median galaxy to median quasar linear sizes is \merr{1.83}{-0.34}{+0.45}
(using $H_0 = 50\,{\rm km\,s^{-1}\,Mpc^{-1}}$, $q_0 = 0.5$; the ratio is
\merr{1.77}{-0.43}{+0.39} using the more up-to-date cosmological parameters
of \citealt{wmap.param}), and the quasar fraction is $0.20 \pm 0.06$.
This lies near the predicted relation for the standard unified scheme,
but also near the curves corresponding to different sets of parameters of
the model of \citet{linsiz.gkkw}.  It is therefore hard to draw a firm
conclusion, but the fact that the MS4 point lies between the points for
the stronger 3CRR sample and those for the weaker 1\,Jy and B3 samples
is consistent with the model of \citet{linsiz.gkkw}.

\section{CONCLUSION}

Studies of powerful, steep-spectrum radio sources have for the last
forty years drawn largely on the 3C sample and its revisions the 3CR
and 3CRR.  It is therefore important to know how representative this
sample is, especially considering the comparative scarcity of radio
sources of high flux density.  An independent southern sample, the
SMS4, has been constructed from the MS4 sample, and compared with 3CRR.
The most important result is that, within the errors, most comparisons
of similarly measured quantities show the properties of the 3CRR and
its southern equivalent, the SMS4, to be very similar.

The MS4 and SMS4 samples have also been used to test some models of
relativistic beaming.  The results appear to be consistent with the model
of \citet{linsiz.gkkw} which predicts that quasar-galaxy differences
are affected by evolution as well as orientation.

\acknowledgments

\section*{Acknowledgements}

We thank staff and students of the School of Physics, University of
Sydney, of Australia Telescope National Facility, and of the
Anglo-Australian Observatory for their generous help with observing and
data reduction.  Special thanks to T. Ye and M. H. Wieringa for their
help with processing and calibrating ATCA data, and to J. C. Baker for
obtaining redshifts for several objects.  Many of the new optical
identifications were made in a night of Director's Time on the
Anglo-Australian Telescope in May 1994.

This research has used data from the COSMOS/UKST Southern Sky Catalogue
provided by the Anglo-Australian Observatory, and the Digitized Sky
Surveys, produced at the Space Telescope Science Institute.  We have
made use of NASA's Astrophysics Data System Abstract Service, and the
NASA/IPAC Extragalactic Database (NED) which is operated by the Jet
Propulsion Laboratory, Caltech, under contract with the US National
Aeronautics and Space Administration.  The Molonglo Observatory
Synthesis telescope is funded by both the Australian Research Council
and the Science Foundation for Physics within the University of Sydney.
The Australia Telescope Compact Array (ATCA) is part of the Australia
telescope, which is funded by the Commonwealth of Australia for operation
as a National Facility managed by CSIRO.  AMB acknowledges the receipt
of an Australian Postgraduate Research Award over the period of this
research.

\clearpage

\addtocounter{figure}{-1}
\begin{figure}[p]
\includegraphics[bb=0 0 643 774,clip,height=627pt,width=442pt]{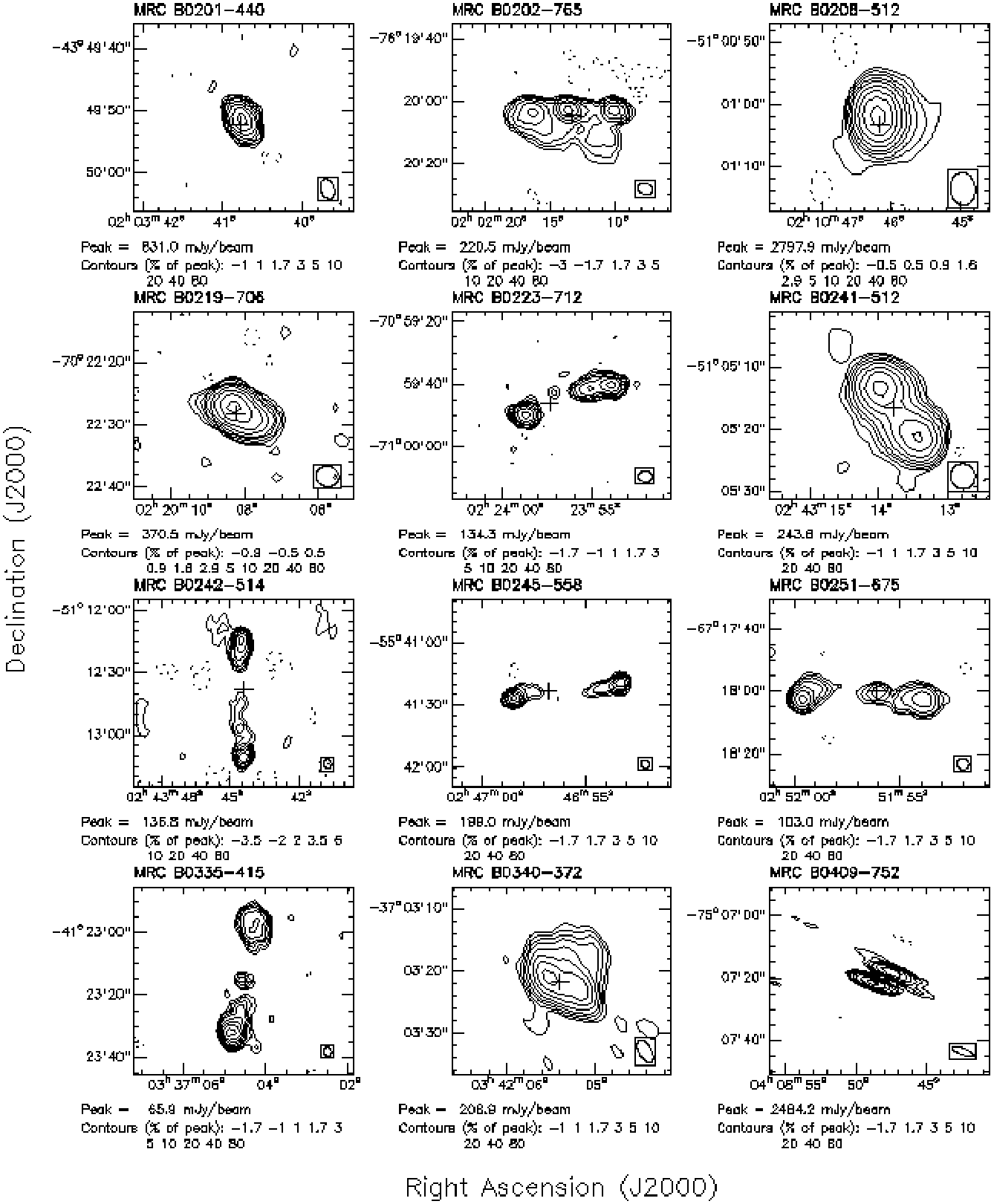}
\caption{ATCA contour maps (continued).}
\end{figure}

\addtocounter{figure}{-1}
\begin{figure}[p]
\includegraphics[bb=0 0 643 774,clip,height=627pt,width=442pt]{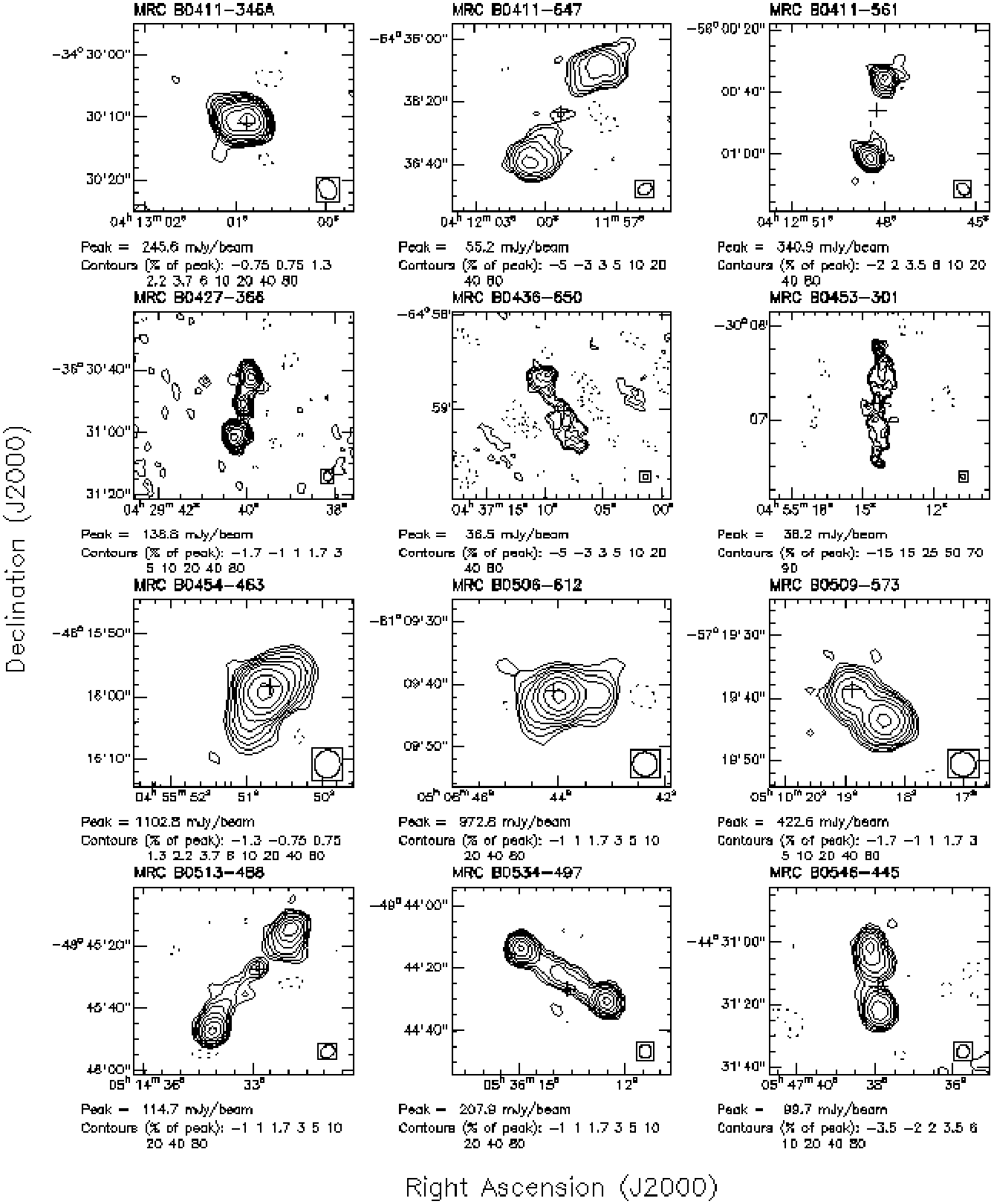}
\caption{ATCA contour maps (continued).}
\end{figure}

\addtocounter{figure}{-1}
\begin{figure}[p]
\includegraphics[bb=0 0 643 774,clip,height=627pt,width=442pt]{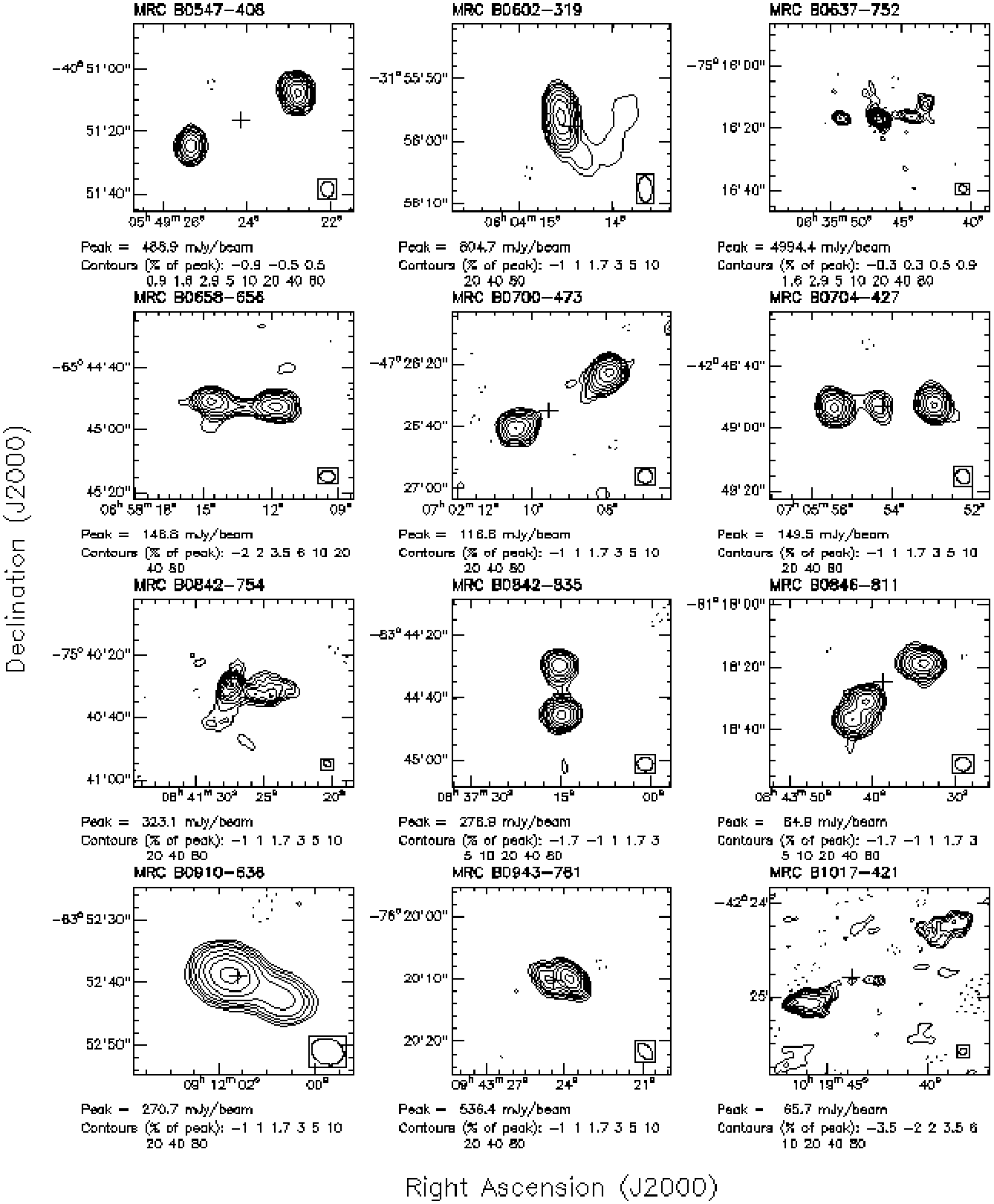}
\caption{ATCA contour maps (continued).}
\end{figure}

\addtocounter{figure}{-1}
\begin{figure}[p]
\includegraphics[bb=0 0 643 774,clip,height=627pt,width=442pt]{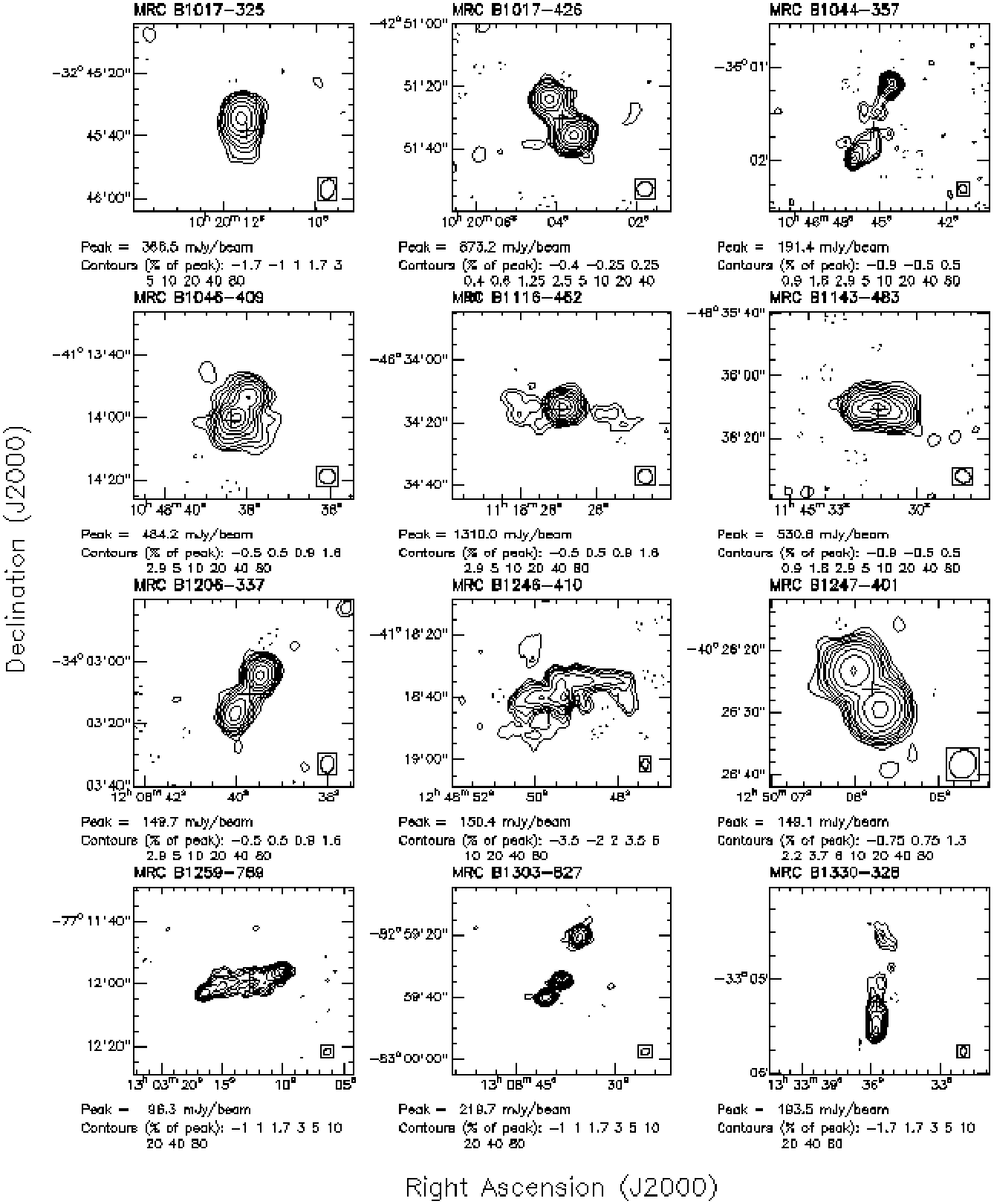}
\caption{ATCA contour maps (continued).}
\end{figure}
\clearpage

\addtocounter{figure}{-1}
\begin{figure}[p]
\includegraphics[bb=0 0 643 774,clip,height=627pt,width=442pt]{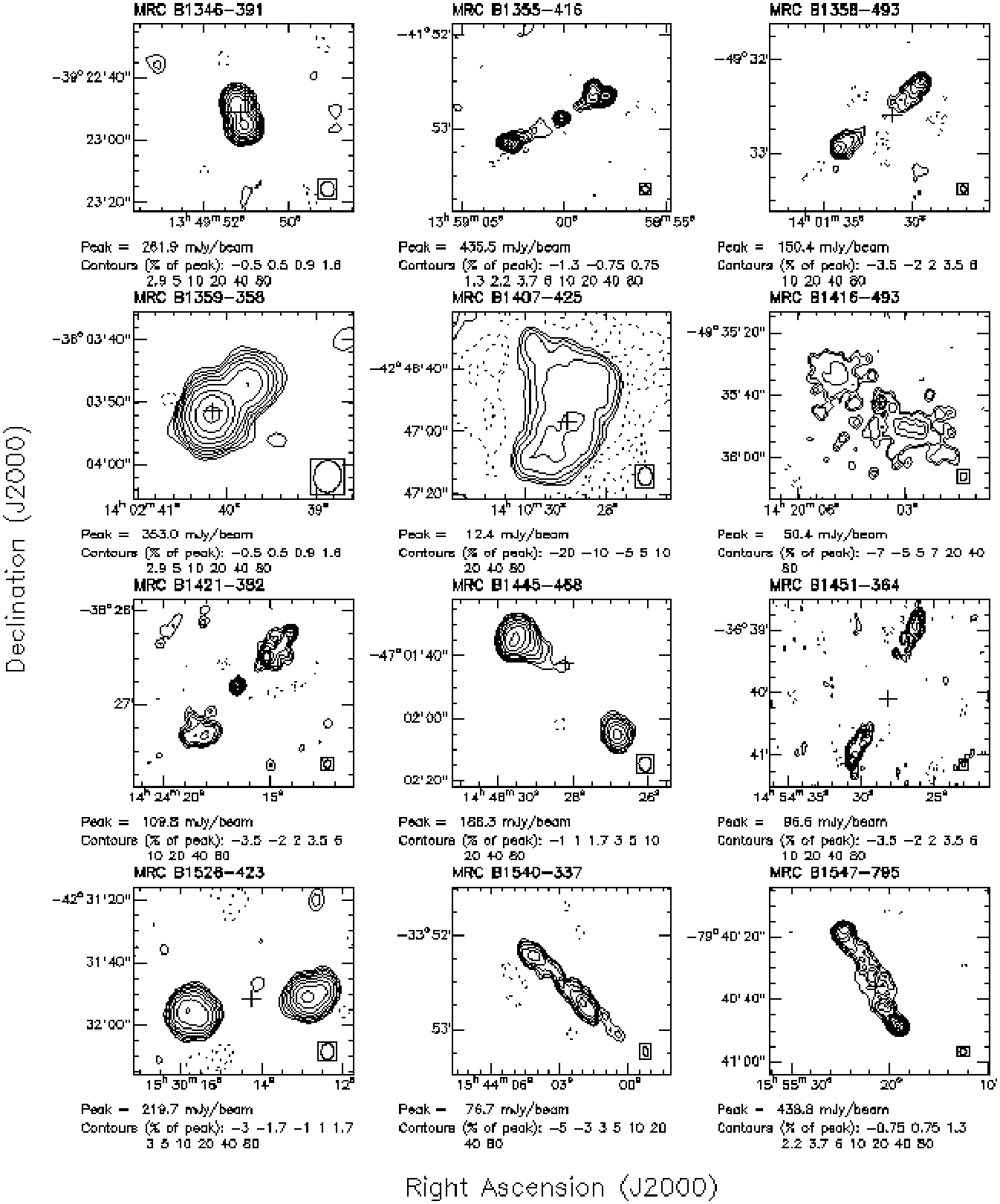}
\caption{ATCA contour maps (continued).}
\end{figure}

\addtocounter{figure}{-1}
\begin{figure}[p]
\includegraphics[bb=0 0 643 774,clip,height=627pt,width=442pt]{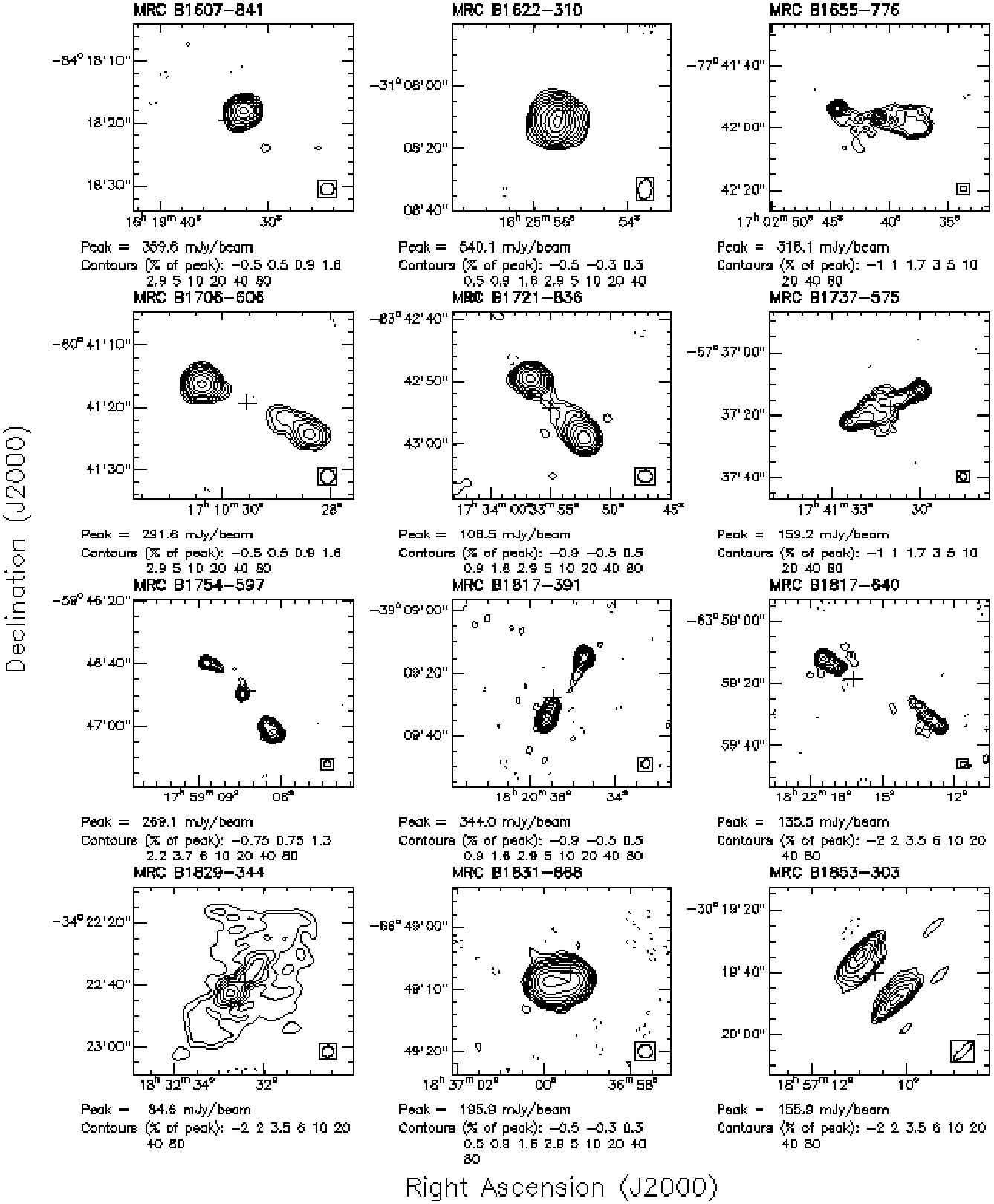}
\caption{ATCA contour maps (continued).}
\end{figure}

\addtocounter{figure}{-1}
\begin{figure}[p]
\includegraphics[bb=0 0 643 774,clip,height=627pt,width=442pt]{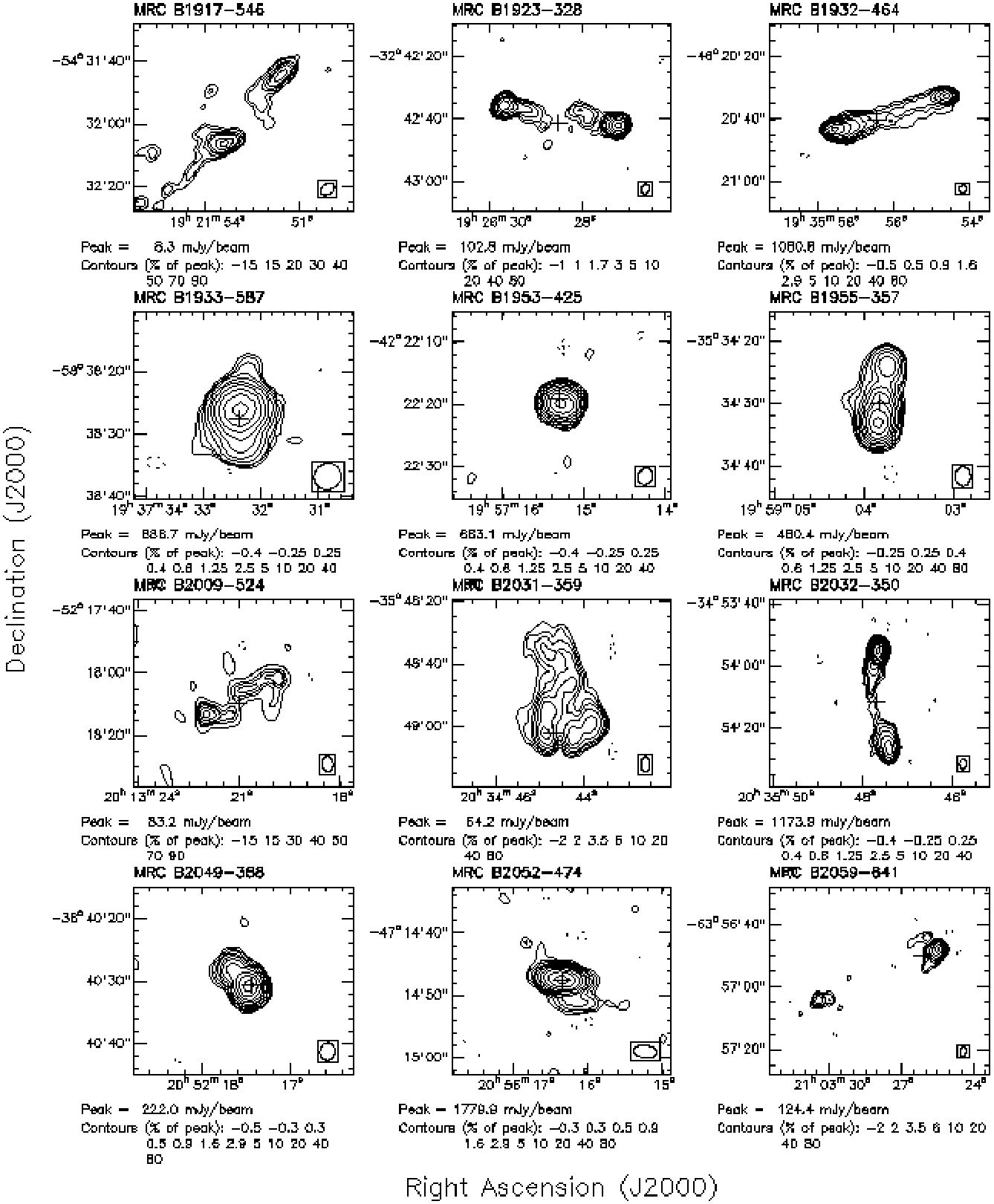}
\caption{ATCA contour maps (continued).}
\end{figure}

\addtocounter{figure}{-1}
\begin{figure}[p]
\includegraphics[bb=0 0 643 774,clip,height=627pt,width=442pt]{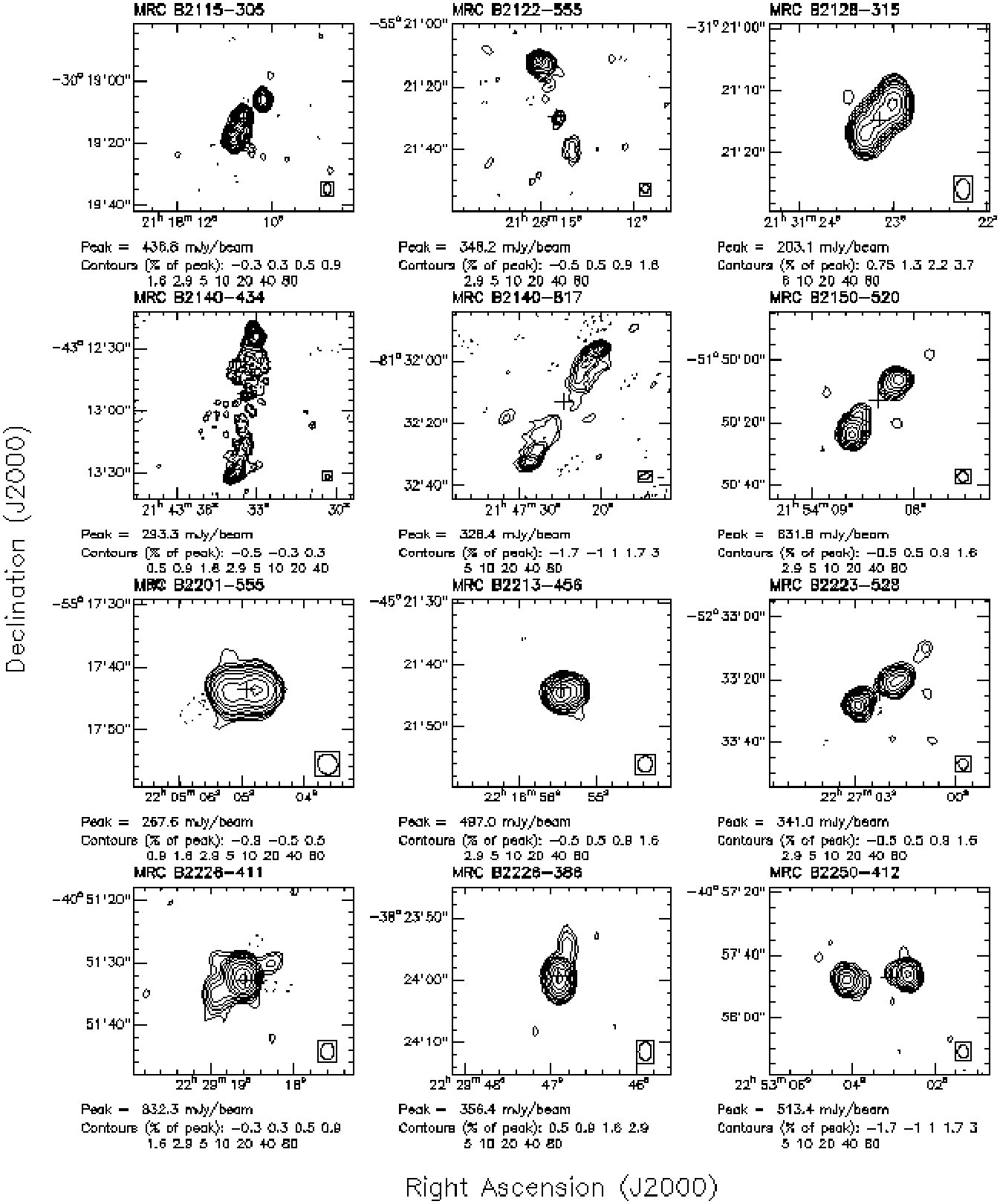}
\caption{ATCA contour maps (continued).}
\end{figure}

\addtocounter{figure}{-1}
\begin{figure}[p]
\includegraphics[bb=0 0 643 774,clip,height=627pt,width=442pt]{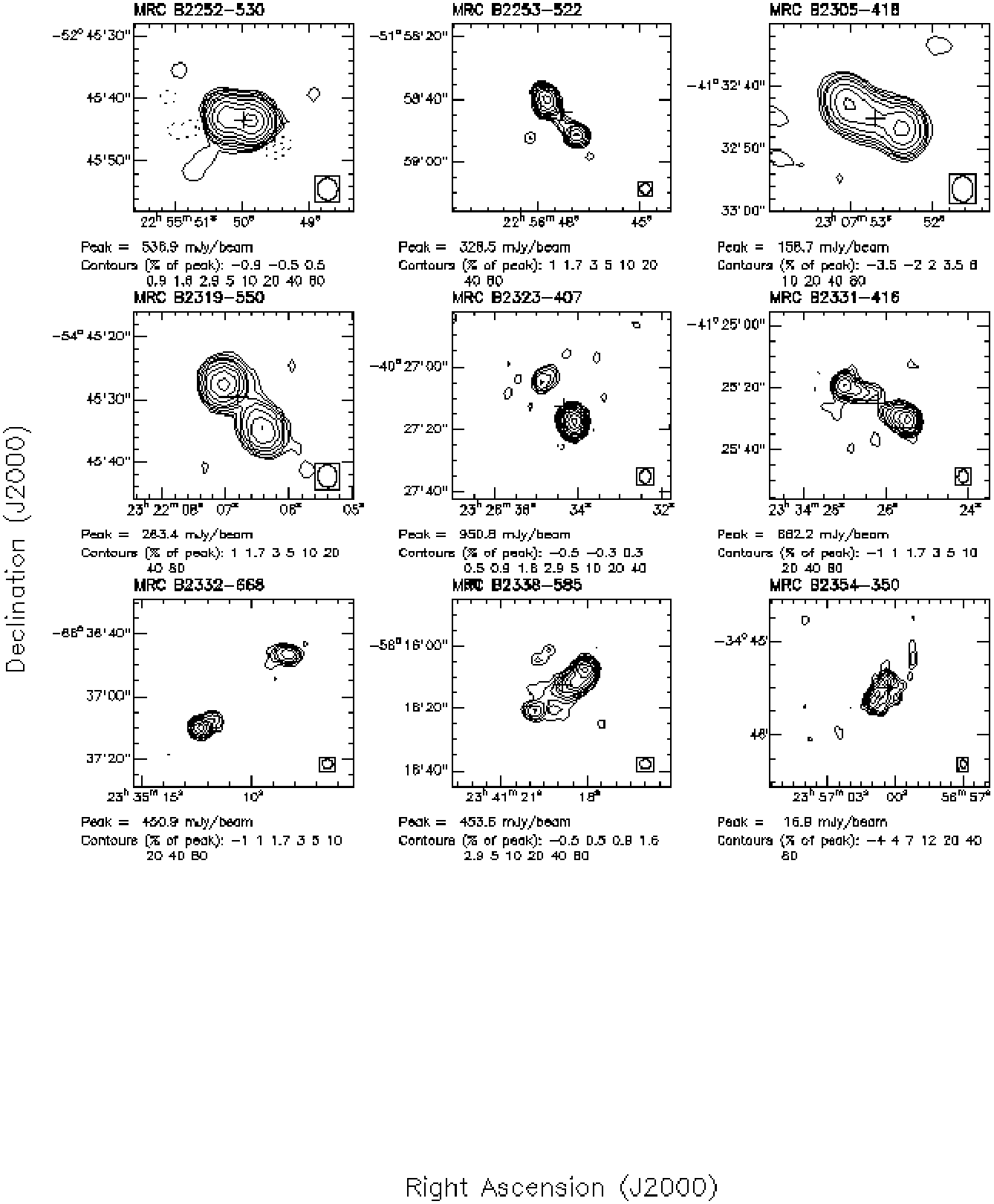}
\caption{ATCA contour maps (continued).}
\end{figure}
\clearpage

\clearpage
\begin{figure} 
\centering
\includegraphics[bb=26 101 563 651,clip,scale=0.4]{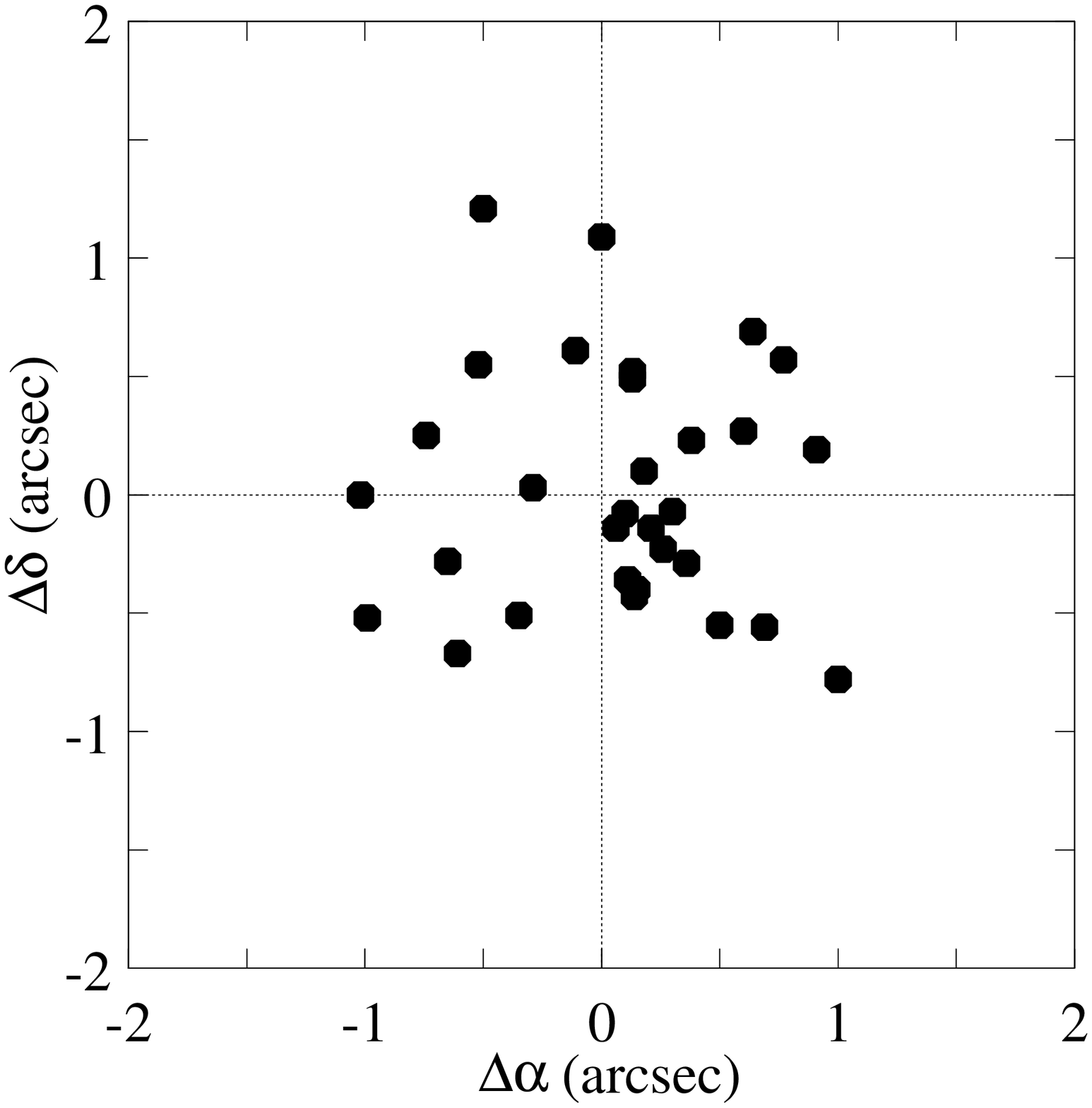}
\caption{Comparison of BP and VLBI positions for 31 sources.
Cross-plot of the difference in declination versus the difference in
right ascension.\label{fig2.radopt}}
\end{figure}

\begin{figure}
\includegraphics[bb=18 20 590 711,clip,scale=0.88]{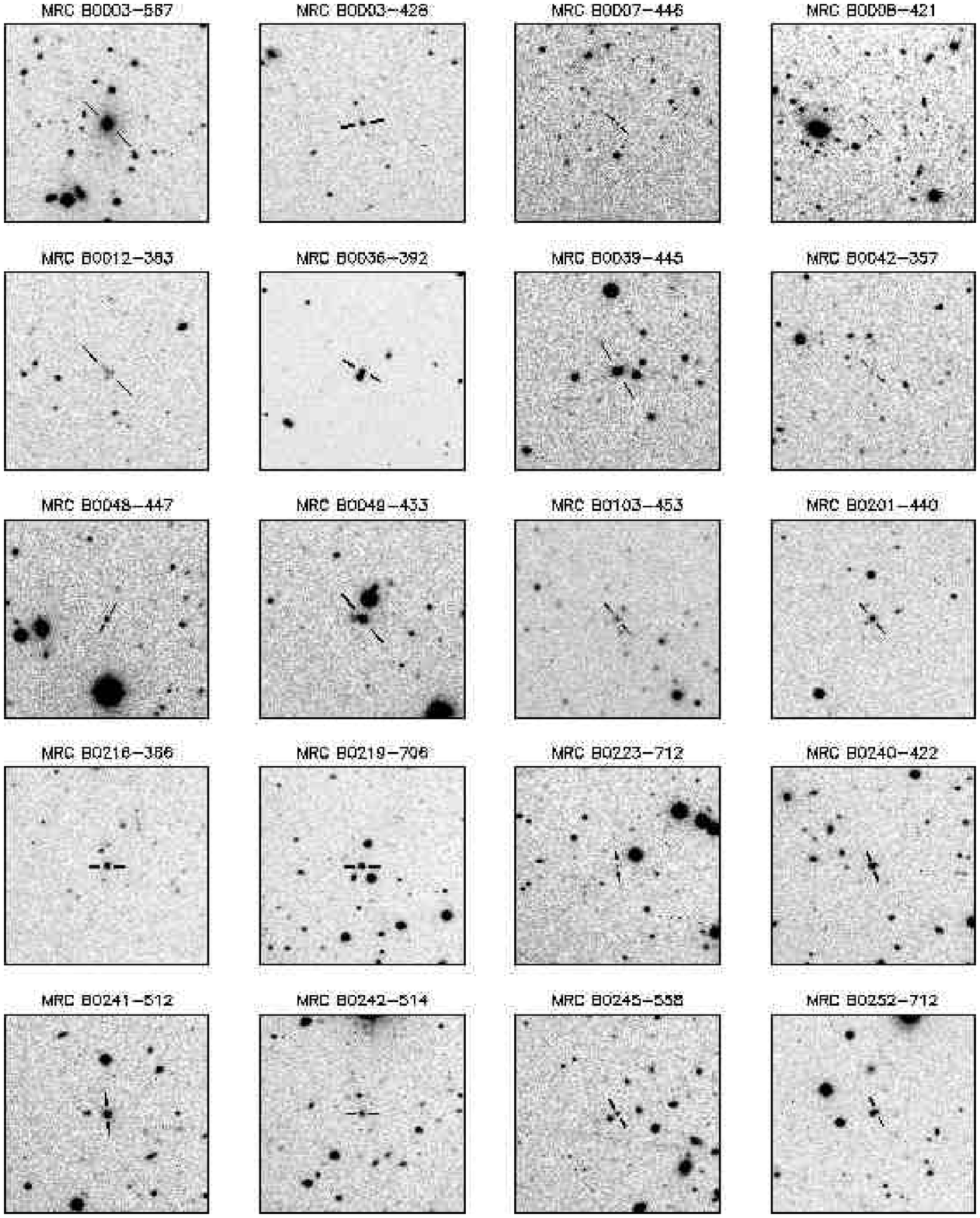}
\caption{AAT CCD images.  The side length of each plot is 1.97$'$.
\label{fig3.ccd}}
\end{figure}

\addtocounter{figure}{-1}
\begin{figure}
\includegraphics[bb=20 20 575 711,clip,scale=0.88]{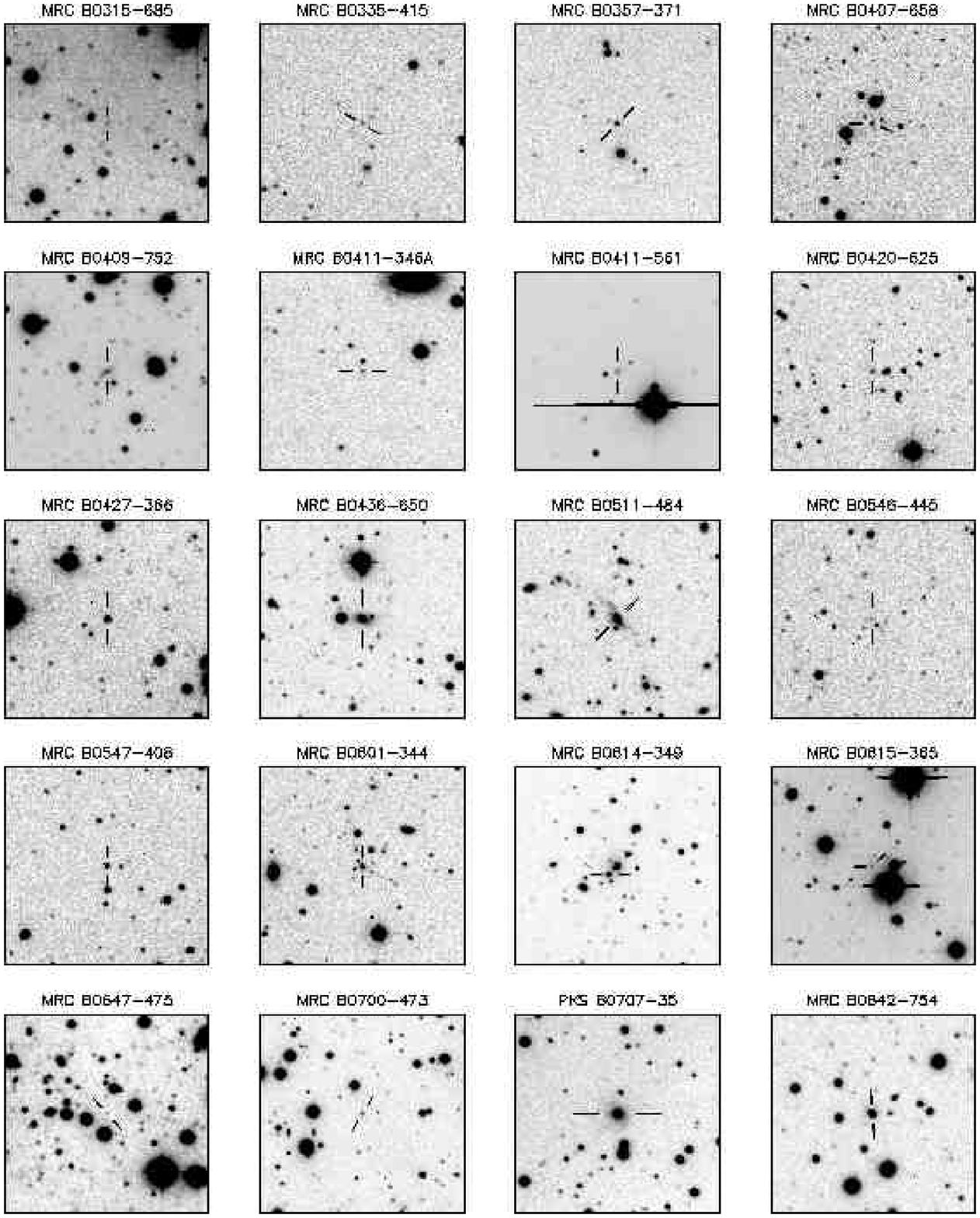}
\caption{AAT CCD images (cont).  The side length of each plot is 1.97$'$.}
\end{figure}

\addtocounter{figure}{-1}
\begin{figure}
\includegraphics[bb=20 20 575 711,clip,scale=0.88]{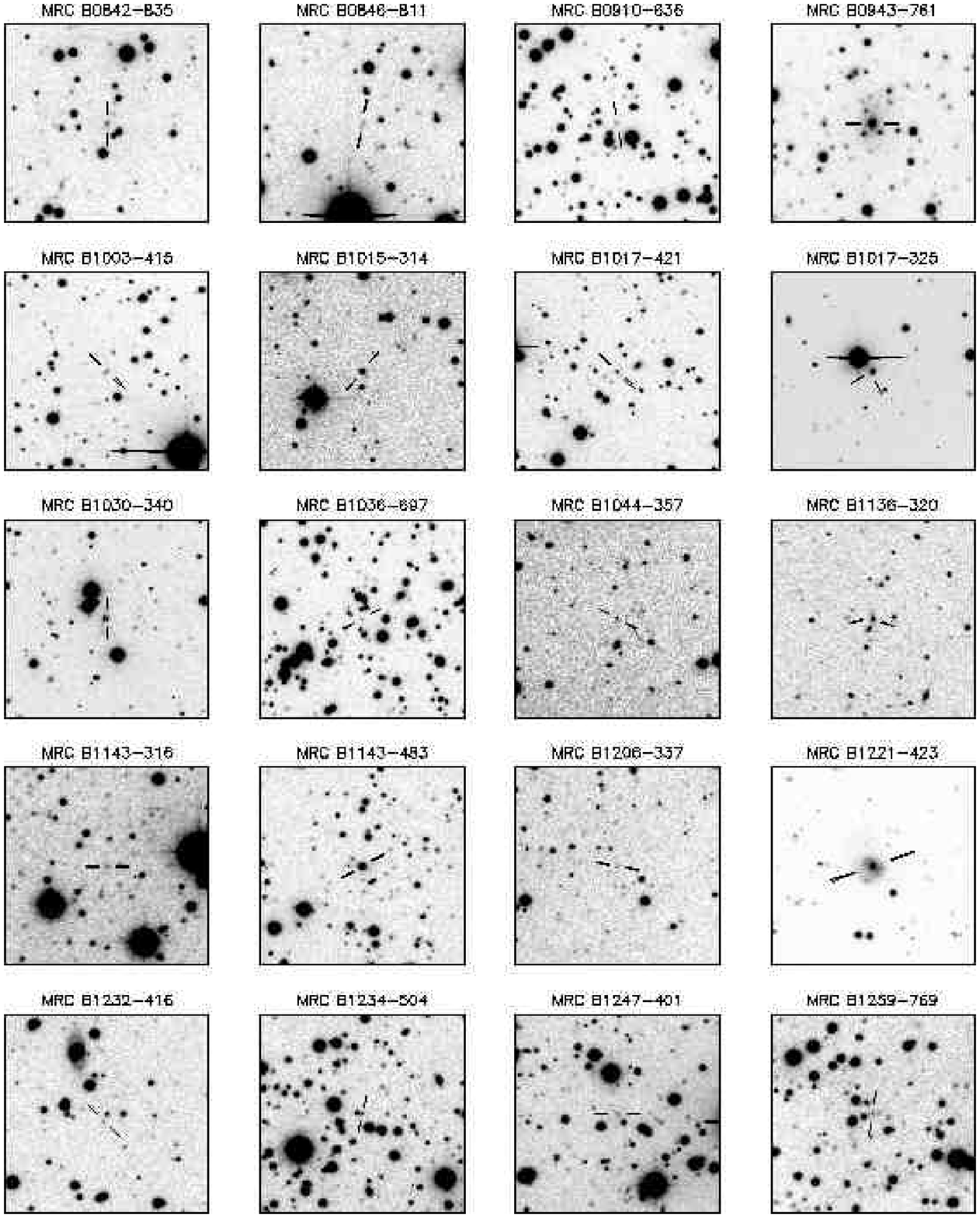}
\caption{AAT CCD images (cont).  The side length of each plot is 1.97$'$.}
\end{figure}

\addtocounter{figure}{-1}
\begin{figure}
\includegraphics[bb=20 20 575 711,clip,scale=0.88]{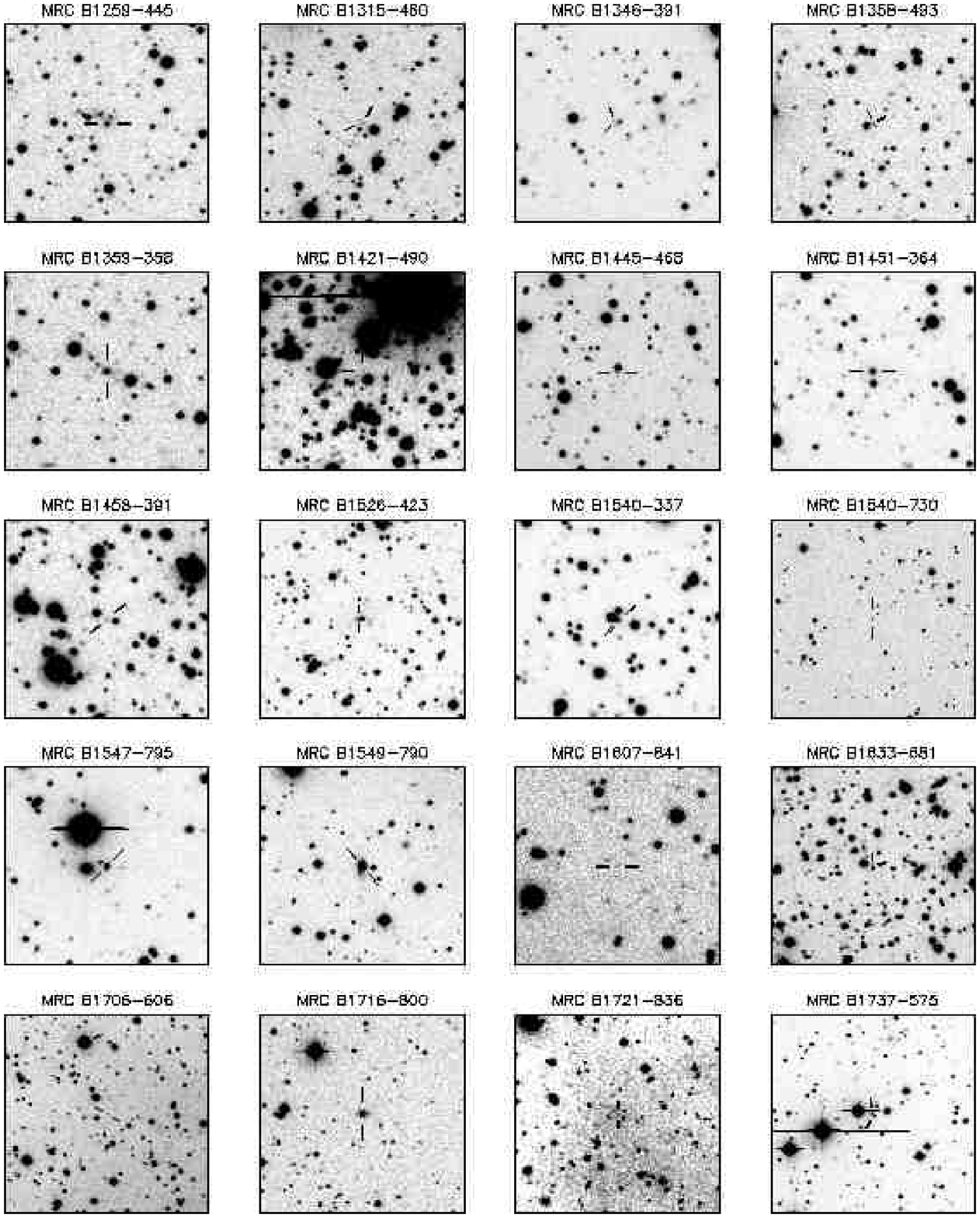}
\caption{AAT CCD images (cont).  The side length of each plot is 1.97$'$.}
\end{figure}

\clearpage

\addtocounter{figure}{-1}
\begin{figure}
\includegraphics[bb=20 20 575 711,clip,scale=0.88]{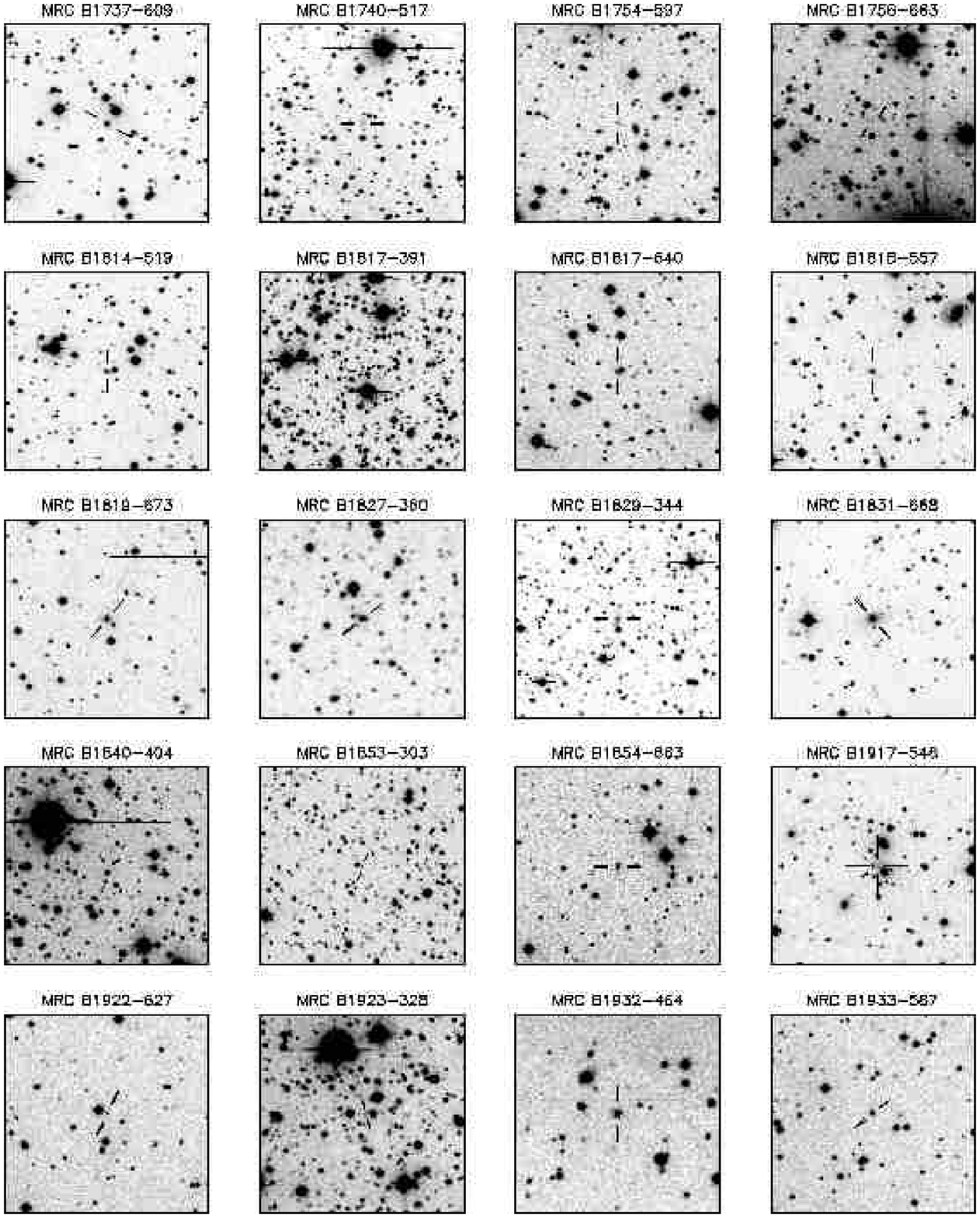}
\caption{AAT CCD images (cont).  The side length of each plot is 1.97$'$.}
\end{figure}

\addtocounter{figure}{-1}
\begin{figure}
\includegraphics[bb=20 20 575 711,clip,scale=0.88]{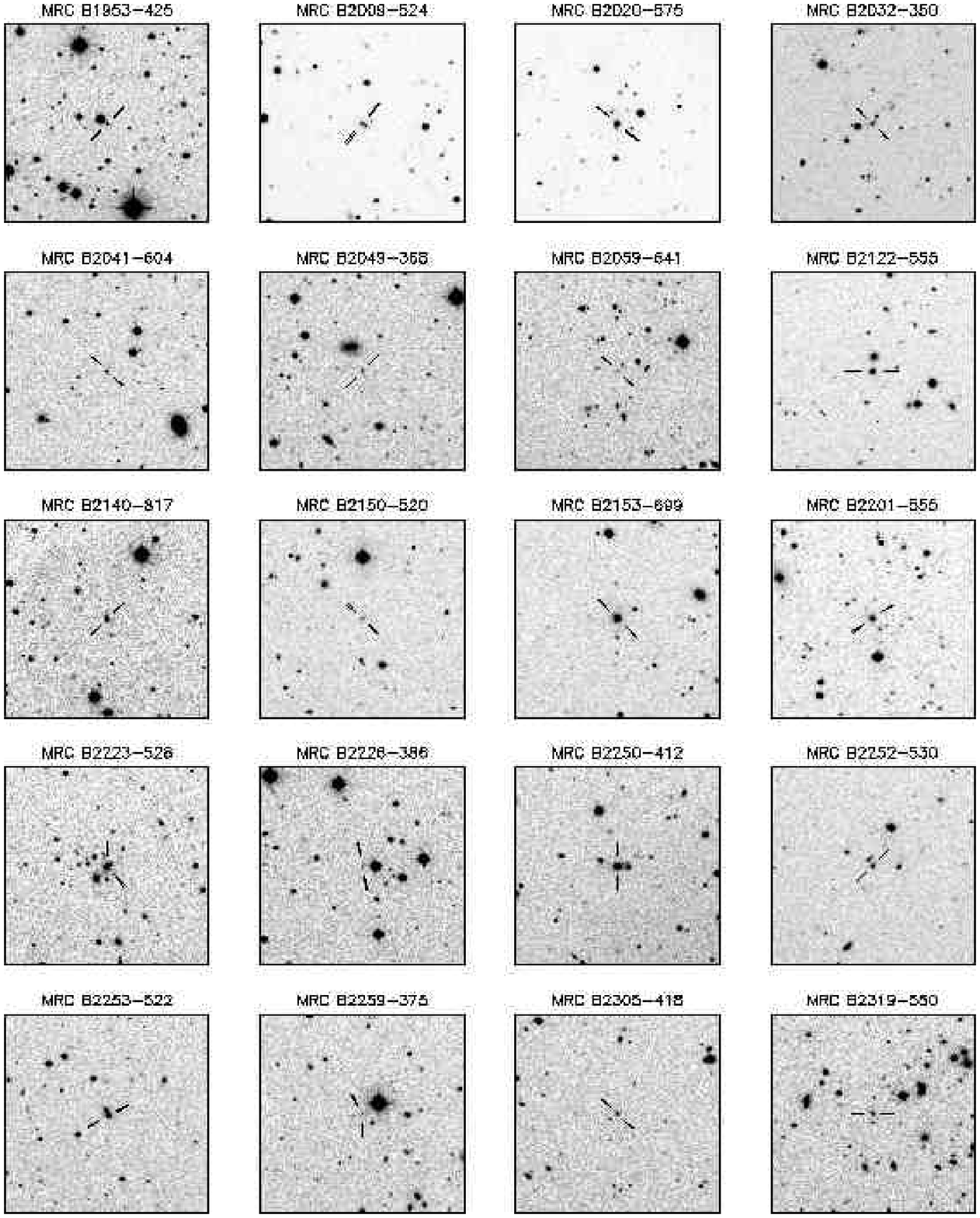}
\caption{AAT CCD images (cont).  The side length of each plot is 1.97$'$.}
\end{figure}

\addtocounter{figure}{-1}
\begin{figure}
\includegraphics[bb=20 20 575 711,clip,scale=0.88]{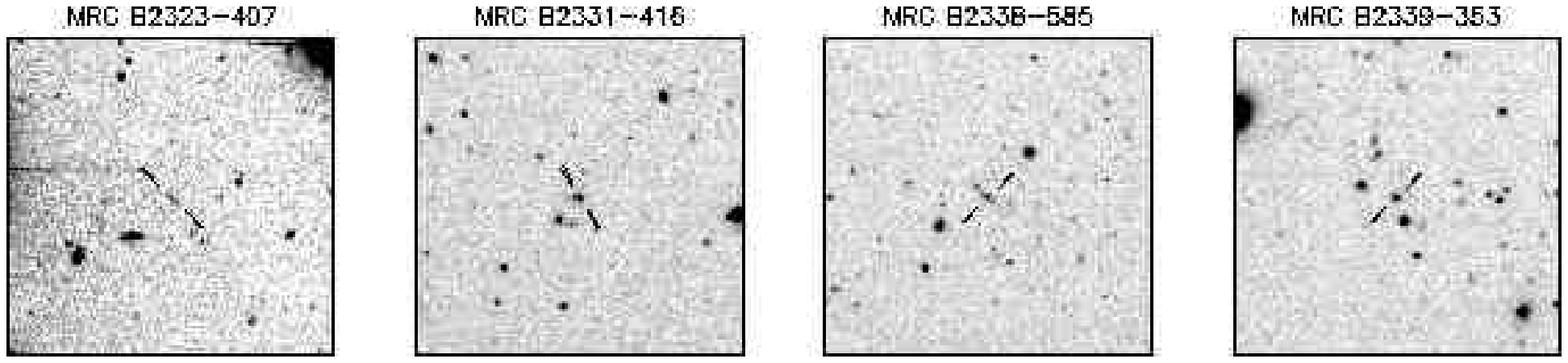}
\caption{AAT CCD images (cont).  The side length of each plot is 1.97$'$.}
\end{figure}

\begin{figure}[htbp]
\centering
\includegraphics[clip,angle=270,scale=0.5]{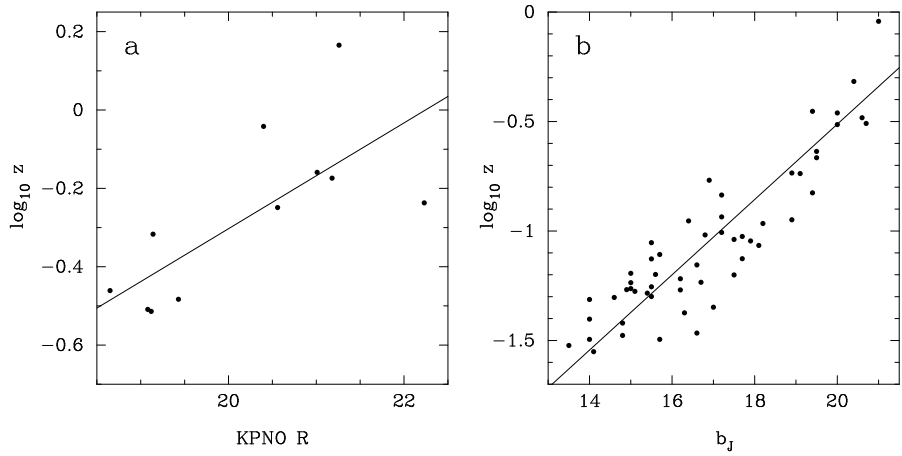}
\caption{(a) Plot of logarithm of redshift versus $R$ magnitude for MS4
galaxies.  The line represents the fit to the data in
Equation~\protect\ref{eq.kpno}.
(b) Plot of logarithm of redshift versus $b_J$ magnitude for MS4
galaxies.  The line represents the fit to the data in
Equation~\protect\ref{eq.bj}.
\label{fig4.magz}}
\end{figure}

\begin{figure}
\centering
\includegraphics[bb=20 100 563 647,clip,scale=0.8]{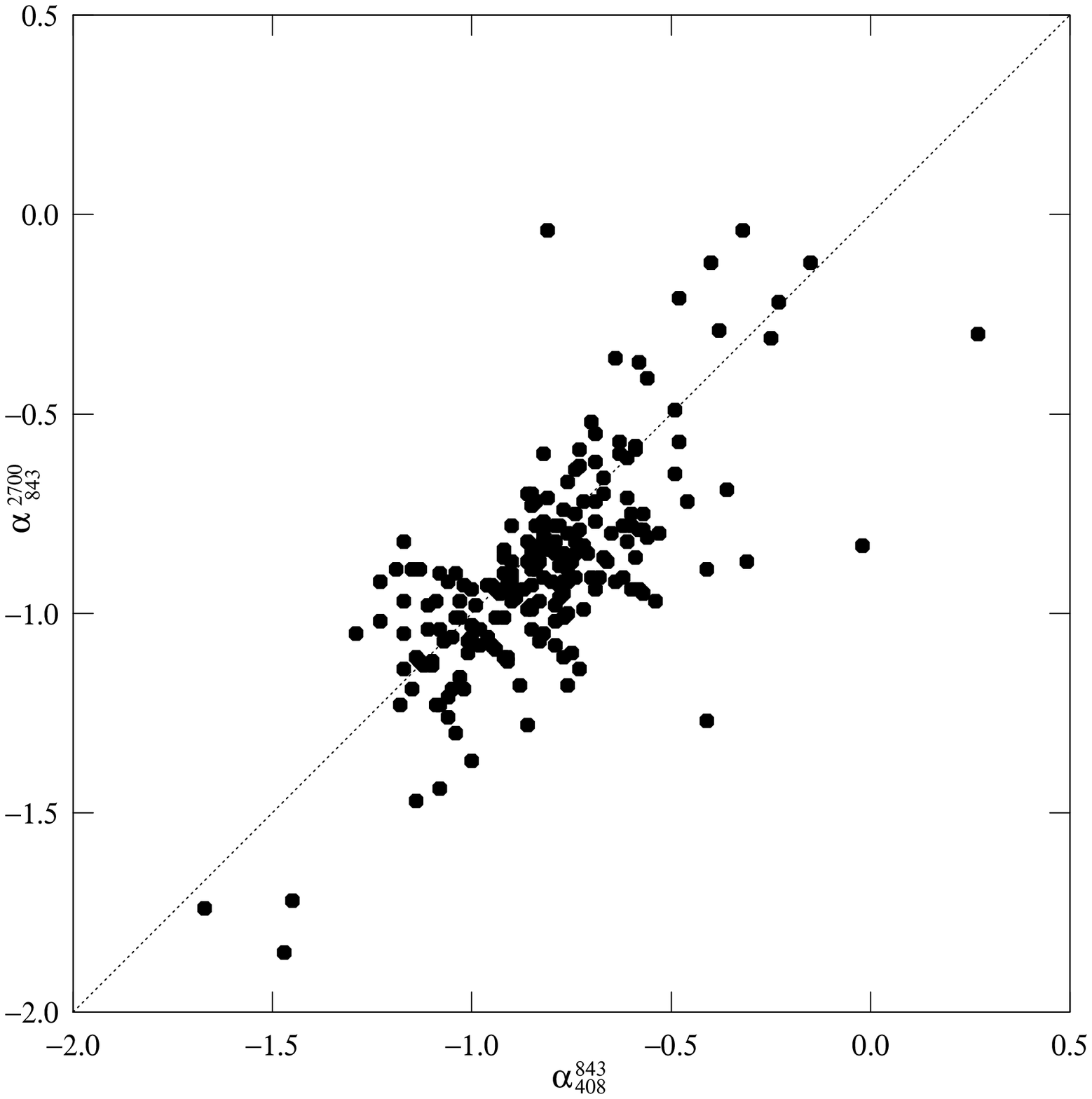}
\caption{Plot of spectral index from 843--2700\,MHz versus spectral
index from 408--843\,MHz for the MS4 sample.  A source with a straight
spectrum ($\alpha_{408}^{843}=\alpha_{843}^{2700}$) would lie somewhere
on the dotted diagonal line.  One source, MRC~B1934$-$638, with
$\alpha_{408}^{843}=1.08$ and $\alpha_{843}^{2700}=-0.16$, lies off the
right edge of the plot.
\label{fig5.col}}
\end{figure}

\begin{figure}
\centering
\includegraphics[bb=156 360 447 672,clip,scale=0.8]{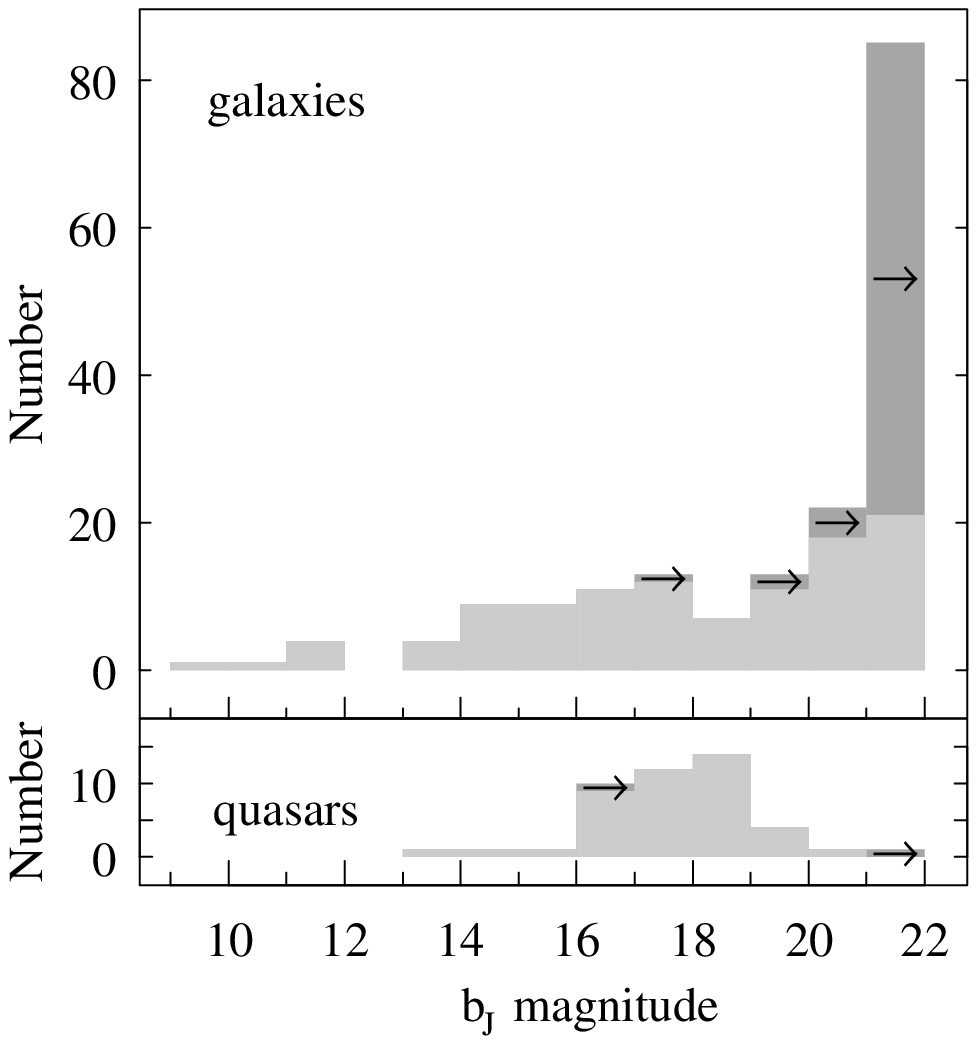}
\caption{Histogram of apparent magnitudes of galaxies (top) and quasars
and quasar candidates (bottom) from the MS4 sample.  The darker grey
bars with arrows represent lower limits to the magnitude for optical
counterparts which are either fainter than the plate limit, or are
confused with other objects.  Most of the galaxies below the limit of
the Schmidt plates were detected on $R$-band CCD images.
\label{fig6.histbj}}
\end{figure}

\clearpage

\begin{deluxetable}{rlcll}
\tabletypesize{\footnotesize}
\tablecaption{Journal of ATCA Observations \label{tab1.atobs}}
\tablehead{
\colhead{(1)} & \colhead{(2)} & \colhead{(3)} & \colhead{(4)} &
\colhead{(5)} \\
\colhead{Run} & \colhead{Date} & \colhead{Frequency} & \colhead{Antenna} & 
\colhead{Secondary Calibrators} \\
\colhead{ } & \colhead{ } & \colhead{(MHz)} & \colhead{Configuration} & 
\colhead{ }}
\startdata
1 & 1990 Aug 24 & 4790 & 
3km \#2\tablenotemark{a} 
& MRC~B2326$-$477, MRC~B2355$-$534 \\
2 & 1990 Aug 25 & 4790 & 
3km \#2\tablenotemark{a}
& MRC~B2311$-$452, MRC~B0008$-$421 \\
3 & 1991 Jan 22 & 4790 & 
3km \#4\tablenotemark{b}
& MRC~B0647$-$475, MRC~B0823$-$500 \\
4 & 1991 Jan 23 & 4790 & 
3km \#4\tablenotemark{b}
& MRC~B1104$-$445, MRC~B1151$-$348 \\
 5 & 1991 Apr 22 & 4740 & 3A & MRC~B0208$-$512, MRC~B0252$-$712 \\
 6 & 1991 Apr 23 & 4740 & 3A & MRC~B1424$-$418\tablenotemark{c}~,
MRC~B1458$-$391 \\
 7 & 1991 Apr 24 & 4740 & 3A & MRC~B1215$-$457, MRC~B1320$-$446 \\
 8 & 1991 Apr 24 & 4740 & 3A & MRC~B0420$-$625, MRC~B0537$-$441 \\
 9 & 1991 Aug 16 & 4740 & 6A & MRC~B1549$-$790,
MRC~B1814$-$637\tablenotemark{c} \\
10 & 1991 Aug 17 & 4740 & 6A & MRC~B1827$-$360, MRC~B1954$-$388 \\
11 & 1992 Jan 16 & 4800 & 6B & MRC~B0022$-$423, MRC~B0153$-$410,
MRC~B0208$-$512 \\
   &          &      &    & MRC~B0332$-$403, MRC~B0438$-$436,
MRC~B1057$-$797 \\
   &          &      &    & MRC~B1234$-$504, MRC~B1458$-$391,
MRC~B1610$-$771 \\
   &          &      &    & PKS~B1934$-$63, MRC~B1954$-$388,
MRC~B2142$-$758 \\
12 & 1992 Apr 17 & 4800 & 6C & MRC~B2106$-$413, MRC~B2149$-$306,
MRC~B2211$-$388 \\
13 & 1993 Dec 19 & 4800 & 6A & PKS~B0642$-$349\tablenotemark{c}~,
MRC~B0646$-$306 \\
\enddata
\tablenotetext{a}{The full name is `3 km minimum redundancy \# 2'.}
\tablenotetext{b}{The full name is `3 km minimum redundancy \# 4'.} 
\tablenotetext{c}{Observed as a calibrator but not subsequently used.}
\end{deluxetable}

\clearpage

\begin{deluxetable}{l@{ }lrrrrc}
\tablecaption{Summary of ATCA observing runs \label{tab2.atobs}}
\tablehead{
\multicolumn{2}{c}{(1)} & \colhead{(2)} & \colhead{(3)} & \colhead{(4)} &
\colhead{(5)} & \colhead{(6)} \\
\multicolumn{2}{c}{Source} & \colhead{Run} & \colhead{No. of} &
\colhead{Int.} & \colhead{Dynamic} & \colhead{Restoring} \\
\multicolumn{2}{c}{Name} & \colhead{No.} & \colhead{cuts} &
\colhead{time} & \colhead{range} & \colhead{beam} \\
\multicolumn{2}{c}{} & \colhead{} & \colhead{} &
\colhead{(min)} & \colhead{} &
\colhead{($\prime\prime \times \prime\prime$, $^{\circ}$)}}
\startdata
MRC & B0003$-$567 & 1 & 7 & 50 & 330 & $3.1 \times 2.6, +25$  \\
MRC & B0003$-$428 & 2 & 7 & 40 & 300 & $3.8 \times 2.8, ~+3$  \\
MRC & B0007$-$446 & 2 & 8 & 51 & 200 & $3.5 \times 2.9, ~+4$  \\
MRC & B0012$-$383 & 2 & 8 & 58 & 540 & $4.3 \times 3.0, ~+1$  \\
MRC & B0013$-$634 & 1 & 8 & 57 & 660 & $3.0 \times 3.0, -90$  \\
 & & & & & &  \\
MRC & B0036$-$392 & 2 & 8 & 49 & 440 & $3.9 \times 3.1, ~+5$  \\
MRC & B0039$-$445 & 2 & 8 & 48 & 530 & $3.6 \times 3.1, ~+7$  \\
\enddata
\end{deluxetable}

\clearpage

\begin{deluxetable}{l@{ }lcllcrrrrr}
\tablecolumns{11}
\tabletypesize{\scriptsize}
\tablecaption{Parameters measured from the ATCA images \label{tab3.atdat}}
\tablehead{
\multicolumn{2}{c}{(1)} & \colhead{(2)} & \colhead{(3)} &
\colhead{(4)} & \colhead{(5)} & \colhead{(6)} & \colhead{(7)} &
\colhead{(8)} & \colhead{(9)} & \colhead{(10)} \\
\multicolumn{2}{c}{Source} &
\colhead{Str.} &
\multicolumn{2}{c}{~R.A.~~(J2000)~~~dec} &
\colhead{pos.} &
\colhead{$S_{\rm int}$} &
\colhead{$S_{\rm core}$} &
\colhead{$R$} &
\colhead{LAS} &
\colhead{PA} \\
\multicolumn{2}{c}{name} &
\colhead{code\tablenotemark{a}} &
\colhead{\llap{h}~~m~~~s} &
\colhead{~$^\circ$~~$'$~~~$''$} &
\colhead{code\tablenotemark{b}} &
\colhead{(mJy)} &
\colhead{(mJy)} &
\colhead{} &
\colhead{($''$)} &
\colhead{($^\circ$)}}
\startdata
MRC & B0003$-$428 & 2 & 00 06 01.99 & $-$42 34 40.8 & ce & 552 & $<$143 & $<$0.35\phn &  5.0 &  62 \\
MRC & B0003$-$567 & 1 & 00 05 57.70 & $-$56 28 31.1 & ce & 635 & \nodata & \nodata & 26.2 & 83 \\
MRC & B0007$-$446 & 2 & 00 10 30.45 & $-$44 22 56.5 & ce & 302 & $<$2 & $<$0.007 & 33.7 &  62 \\
MRC & B0012$-$383 & 2 & 00 15 24.37 & $-$38 04 35.0 & ce & 307 & $<$55 & $<$0.22\phn & 25.3 &  58 \\
MRC & B0013$-$634 & 2 & 00 16 02.89 & $-$63 10 04.8 & co & 417 & 8\rlap{:} & 0.021: & 29.4 &  76 \\
 & & & & & & & & & & \\
MRC & B0036$-$392 & D2 & 00 38 26.84 & $-$38 59 48.8 & co & 454 & 177\rlap{:} & 0.64:\phn & 4.6 & 40 \\
MRC & B0039$-$445 & 2 & 00 42 08.96 & $-$44 13 59.8 & ce & 1134 & $<$204 & $<$0.22\phn &  6.2 & $-$79 \\
\enddata
\tablenotetext{a}{Code for source structure.  ``1'' = FR\,1
double, ``2'' = FR\,2 double, ``1/2'' = structure
borderline between FR\,1 and FR\,2, ``D2'' = double with one component
coincident with the optical counterpart \citep{d2.ref}, ``T'' =
core-dominated triple, ``c'' = structure not resolved well enough to classify,
``x'' = extended source which does not fall into any of the usual categories.}
\tablenotetext{b}{Code for the type of position given in columns 2 and 3.
``p'' stands for peak, ``ce'' for centroid, ``co'' for core, ``f'' for the
peak of a Gaussian fit to a compact source, and ``m'' for the midpoint
of the two inner peaks of a tailed radio source.  A colon ``:'' after the
position code denotes that the position is suspect, usually because
the source was observed in Run~9: see Section~\ref{sec.atpos}.}
\tablenotetext{c}{Position is from the MOST calibrator list.  The source was
a secondary calibrator observed before a good ATCA position was available.}
\end{deluxetable}

\clearpage

\begin{deluxetable}{crrr}
\tablecaption{Mean and median separations from candidate to position of
confirmed optical counterpart for 43 FR\,2 sources.  Separations are in
units of largest angular extent.  Here ``midway point'' stands for the
point midway between the midpoint and the centroid, and $\sigma$ is the
standard deviation of the mean.  \label{tab4.radopt}}
\tablehead{
\colhead{Radio search} & \multicolumn{3}{c}{Radio-optical Separation} \\
\cline{2-4}
\colhead{position} & \colhead{Median} & \colhead{Mean} & \colhead{$\sigma$}}
\startdata
    Centroid & 0.081 & 0.101 & 0.077 \\
    Midpoint & 0.078 & 0.093 & 0.069 \\
Midway point & 0.057 & 0.074 & 0.057 \\
\enddata
\end{deluxetable}

\clearpage

\begin{deluxetable}{clccl}
\tablecaption{Journal of AAT imaging observations \label{tab5.aatobs}}
\tablehead{
\colhead{Run} & \colhead{Date} & \colhead{Observers} & \colhead{Seeing} & 
\colhead{Remarks} \\
\colhead{No.} & \colhead{} & \colhead{} & \colhead{(arcsec)} & 
\colhead{}}
\startdata
1 & 1993 Feb 26 & Burgess, Hunstead & 1.6 & Mostly clear \\
2 & 1993 Feb 27 & '' & 1.6 & Clear \\
3 & 1993 Aug 15 & Burgess, Sadler & 1.4 & Mostly cloudy \\
4 & 1993 Aug 16 & '' & 2.1 & Partly cloudy \\
5 & 1994 Feb 14 & Da Costa, Heisler & 1.6 & Service run \\
6 & 1994 May 14 & Baker, Hunstead, Burgess & 1.2 & Clear \\
\enddata
\end{deluxetable}

\clearpage

\begin{deluxetable}{cl}
\tablecaption{Photometric solutions for AAT CCD imaging.
\label{tab6.phot}}
\tablehead{
\colhead{Runs} & \colhead{Photometric solution\tablenotemark{a}}}
\startdata
1, 2 & $R = 31.03 - 2.5\,\log_{10} (ADU s^{-1}) - 0.205\,\chi$ \\
6    & $R = 30.97 - 2.5\,\log_{10} (ADU s^{-1}) - 0.109\,\chi$ \\
\enddata
\tablenotetext{a}{$ADU s^{-1}$ stands for the number of counts in
Analogue to Digital Units per second, and $\chi$ stands for air-mass.}
\end{deluxetable}

\clearpage

\begin{deluxetable}{l@{ }lcllrrclrrrrr}
\tablecolumns{14}
\tabletypesize{\footnotesize}
\rotate
\tablewidth{600pt}
\tablecaption{Positions, magnitudes, and redshifts for the optical
counterparts.  \label{tab7.opt}}
\tablehead{
\multicolumn{2}{c}{(1)} & \colhead{(2)} & \colhead{(3)} & \colhead{(4)} &
\colhead{(5)} & \colhead{(6)} & \colhead{(7)} & \colhead{(8)} &
\multicolumn{5}{c}{(9)} \\
\multicolumn{2}{c}{Source} &
\colhead{ID\tablenotemark{a}} &
\multicolumn{2}{c}{Optical position} &
\colhead{$b_J$} &
\colhead{$R$} &
\colhead{Seeing} &
\colhead{$z$} &
\multicolumn{5}{c}{References} \\
\multicolumn{2}{c}{Name} &
\colhead{} &
\multicolumn{2}{c}{R. A.~~(J2000)~~dec} &
\colhead{mag} &
\colhead{mag} &
\colhead{($''$)} &
\colhead{} &
\colhead{ID} & \colhead{FC} & \colhead{pos} & \colhead{$b_J$} & \colhead{$z$}
}
\startdata
MRC & B0003$-$567 & g & 00 05 57.94 & $-$56 28 31.5 & 19.6 & 18.7 & 1.4 & \llap{(}0.26) & 1 & 1 & 2 & 3 & 4 \\
MRC & B0003$-$428 & g & 00 06 01.76 & $-$42 34 41.4 & 21.2 & 20.2 & 1.5 & \llap{(}0.53) & 5 & 5 & 2 & 3 & 6 \\
MRC & B0003$-$833 & g & 00 06 16.64 & $-$83 06 07.8 & 20.1 & \nodata & \nodata & \llap{(}0.32) & 7 & 5 & 7 & 3 & 4 \\
MRC & B0007$-$446 & g & 00 10 29.89 & $-$44 22 55.4 & \llap{$>$}21.5\tablenotemark{b} & 22.2 & 1.3 & \llap{(}1.00) & 8 & 8 & 9 & \nodata & 6 \\
MRC & B0008$-$421 & g & 00 10 52.54 & $-$41 53 10.6 & \llap{$>$}21.5\tablenotemark{b} & 22.6 & 1.3 & \llap{(}1.12) & 10 & 11 & 12 & \nodata & 6 \\
\\
MRC & B0012$-$383 & g & 00 15 24.31 & $-$38 04 38.8 & \llap{$>$}21.5\tablenotemark{b} & 20.4 & 1.4 & \llap{(}0.57) & 5 & 5 & 9 & \nodata & 6 \\
MRC & B0013$-$634 & g & 00 16 02.82 & $-$63 10 04.8 & \llap{$>$}21.5 & \nodata & \nodata & \llap{($>$}0.56) & 5 & \nodata & 2 & \nodata & 4 \\
\enddata
\tablenotetext{a}{\scriptsize Type of optical ID: g = galaxy, g? = possible galaxy,
l = BL~Lac object, Q = quasar, Q? = quasar candidate,
BF = blank field, O = field obscured by star,
R = unidentified relic radio source in cluster}
\tablenotetext{b}{\scriptsize Blank field on the Schmidt plates}
\tablenotetext{c}{\scriptsize Crowded field not well analysed by COSMOS}
\tablenotetext{d}{\scriptsize Extended radio source with more than one candidate}
\tablenotetext{e}{\scriptsize Compact radio source with large radio-optical offset}
\tablenotetext{f}{\scriptsize Position quoted is that of the VLA core}
\tablenotetext{g}{\scriptsize The redshift quoted is that of the galaxy NGC~5419}
\tablenotetext{h}{\scriptsize The redshift quoted is that of the cluster Abell~3667}
\end{deluxetable}

\addtocounter{table}{-1}
\begin{deluxetable}{l@{ }lcllrrclrrrrr}
\tablecolumns{14}
\tabletypesize{\footnotesize}
\rotate
\tablewidth{600pt}
\tablecaption{Positions, magnitudes, and redshifts for the optical
counterparts.}
\tablehead{
\multicolumn{2}{c}{(1)} & \colhead{(2)} & \colhead{(3)} & \colhead{(4)} &
\colhead{(5)} & \colhead{(6)} & \colhead{(7)} & \colhead{(8)} &
\multicolumn{5}{c}{(9)} \\
\multicolumn{2}{c}{Source} &
\colhead{ID\tablenotemark{a}} &
\multicolumn{2}{c}{Optical position} &
\colhead{$b_J$} &
\colhead{$R$} &
\colhead{Seeing} &
\colhead{$z$} &
\multicolumn{5}{c}{References} \\
\multicolumn{2}{c}{Name} &
\colhead{} &
\multicolumn{2}{c}{R. A.~~(J2000)~~dec} &
\colhead{mag} &
\colhead{mag} &
\colhead{($''$)} &
\colhead{} &
\colhead{ID} & \colhead{FC} & \colhead{pos} & \colhead{$b_J$} & \colhead{$z$}
}
\startdata
 &  &  &  &  &  &  &  &  &  &  &  & \\
\enddata
\end{deluxetable}

\clearpage

\begin{deluxetable}{rrcllllcl}
\tablecaption{Sample size and median values of various
quantities for sources in the MS4, SMS4, and 3CRR samples.
\label{tab8.medians}}
\tablehead{
\colhead{(1)} & \colhead{(2)} & \colhead{(3)} &
\colhead{(4)} & \colhead{(5)} & \colhead{(6)}\\
\colhead{} & \colhead{} & \multicolumn{4}{c}{Median values} \\
\cline{3-6}
\colhead{Sample} & \colhead{N} & \colhead{LAS} &
\colhead{$z$} & \colhead{$l$} & \colhead{$P_{408}$} \\
\colhead{} & \colhead{} & \colhead{($''$)} &
\colhead{} & \colhead{(kpc)} & \colhead{(W\,Hz$^{\rm -1}$)}}
\startdata
& & & & & \\
MS4 & 228 & \merr{27}{-2}{+5} & \merr{0.53}{-0.05}{+0.05} & \merr{126}{-9}{+22} & \merr{5.9}{-0.9}{+2.2} $\times 10^{27}$ \\
 &  &  &  &  &  \\
SMS4 & 137 & \merr{34}{-2}{+6} & \merr{0.50}{-0.07}{+0.06} & \merr{159}{-11}{+45} & \merr{6.6}{-1.8}{+5.7} $\times 10^{27}$ \\
 &  &  &  &  &  \\
3CRR & 172 & \merr{32}{-6}{+11} & \merr{0.50}{-0.08}{+0.07} & \merr{137}{-33}{+33} & \merr{8.4}{-3.2}{+4.0} $\times 10^{27}$ \\
 &  &  &  &  &  \\
& & & & & \\
\enddata
\end{deluxetable}

\clearpage

\begin{deluxetable}{rrrcrrcrr}
\tablecaption{Numbers and percentages of galaxies and quasars in the
MS4, SMS4, and 3CRR samples.  \label{tab9.ngq}}
\tablehead{
\colhead{Sample} & \multicolumn{2}{c}{Galaxies} & &
\multicolumn{2}{c}{Quasars} & &
\multicolumn{2}{c}{Unclassified} \\
\cline{2-3} \cline{5-6} \cline{8-9}
\colhead{} & \colhead{N} & \colhead{\%} & &
\colhead{N} & \colhead{\%} & &
\colhead{N} & \colhead{\%}}
\startdata
 MS4 & 174 & 76.3 & & 39 & 17.1 & & 15 & 6.6 \\
SMS4 & 109 & 79.6 & & 19 & 13.9 & & 9 & 6.6 \\
3CRR & 129 & 75.0 & & 43 & 25.0 & & 0 & 0.0 \\
\enddata
\end{deluxetable}

\clearpage

\tabletypesize{\footnotesize}
\begin{deluxetable}{l@{ }lcrrr}
\tablecolumns{6}
\tablewidth{0pt}
\tablecaption{Radio-structure data from the literature. \label{tab10.struc}}
\tablehead{
\multicolumn{2}{c}{(1)} & \colhead{(2)} & \colhead{(3)} &
\colhead{(4)} & \colhead{(5)} \\
\multicolumn{2}{c}{Source} & \colhead{Struc.} & \colhead{LAS} & \colhead{PA} &
\colhead{Map} \\
\multicolumn{2}{c}{name} & \colhead{code\tablenotemark{a}} &
\colhead{($''$)} & \colhead{(deg)} & \colhead{ref.}
}
\startdata
MRC & B0003$-$833 & 2 & 69\phd & $-$26 & 1 \\
MRC & B0023$-$333 & 1 & 68\phd & 17 & 2 \\
MRC & B0039$-$445 & 2 & 6.5 & $-$79 & 3 \\
MRC & B0043$-$424 & 2 & 144\phd & $-$45 & 3 \\
MRC & B0103$-$453 & 2 & 140\phd & 54 & 2 \\
 & & & & & \\
MRC & B0110$-$692 & 2 & 77\phd & $-$13 & 1 \\
PKS & B0114$-$47 & 2 & 690\phd & $-$18 & 4 \\
\enddata
\tablenotetext{a}{Code for source structure.  ``1'' = FR\,1
double, ``2'' = FR\,2 double, ``1/2'' = structure
borderline between FR\,1 and FR\,2, ``D2'' = double with one component
coincident with the optical counterpart \citep{d2.ref}, ``T'' =
core-dominated triple, ``c'' = structure not resolved well enough to classify,
``x'' = extended source which does not fall into any of the usual categories.}
\tablerefs{
(1) P. A. Jones, unpublished ATCA image (1992); 
(2) MOST image, Paper I; 
(3) \citet{newmaps.2jy}; 
(4) \citet{rad.S16}; 
(5) \citet{map.morganti}; 
(6) MOST image, \citet{rad.J1}; 
(7) \citet{opt.k2}; 
(8) \citet{rad.S14}; 
(9) R. Subrahmanyan, unpublished ATCA image (1992); 
(10) \citet{map.0409}; 
(11) \citet{opt.r1}; 
(12) \citet{rad.C5}; 
(13) \citet{0521.radxray}; 
(14) \citet{map.parma}; 
(15) \citet{0743.rayner}; 
(16) \citet{map.sproats}; 
(17) \citet{opt.s38}; 
(18) \citet{map.fleurs1415}; 
(19) \citet{rad.M3}; 
(20) \citet{rad.E1}; 
(21) M. Wieringa, unpublished ATCA image; 
(22) \citet{1932.jetcloud}; 
(23) R. Hunstead, unpublished ATCA image; 
(24) \citet{fosb.2152}; 
(25) \citet{2250-412.radoptx}. 
}
\end{deluxetable}

\clearpage

\begin{deluxetable}{rrrrrrr}
\tablecaption{Numbers and percentages of
Fanaroff-Riley class 1 and 2 sources in the MS4, SMS4, and 3CRR
samples.  Since not all MS4 sources have been imaged at high enough
resolution, the structures of 47  sources are uncertain.
The structures of 5
MS4 sources do not fit into either FR\,1 or FR\,2 categories.
The 3CRR classifications are from the compilation of Laing (1998,
priv. comm.).  \label{tab11.fr}}
\tablehead{
\colhead{Sample} &
\multicolumn{2}{c}{FR\,1} & \multicolumn{2}{c}{FR\,2} &
\multicolumn{2}{c}{Unclassified} \\
\colhead{} & \colhead{N} & \colhead{\%} & \colhead{N} & \colhead{\%} &
\colhead{N} & \colhead{\phantom{000}\%}}
\startdata
 MS4 & 34 & 14.9 & 142 & 62.3 & 52 & 22.8 \\
SMS4 & 22 & 16.1 & 91 & 66.4 & 24 & 17.5 \\
3CRR & 28 & 16.3 & 141 & 82.0 & 3 & 1.7 \\
\enddata
\end{deluxetable}

\clearpage

\begin{deluxetable}{crl}
\tablecaption{Median redshifts for confirmed quasars in the MS4, SMS4, and
3CRR samples.  \label{tab12.medz}}
\tablehead{
\colhead{Sample} & \colhead{$n$} & \colhead{Median} \\
\colhead{} & \colhead{} & \colhead{quasar $z$}}
\startdata
MS4 & 39 & \merr{0.87}{-0.19}{+0.13} \\
 &  &  \\
SMS4 & 19 & \merr{0.65}{-0.12}{+0.22} \\
 &  &  \\
3CRR & 43 & \merr{0.90}{-0.06}{+0.14} \\
\enddata
\end{deluxetable}

\clearpage

\begin{deluxetable}{crrrcrrrcrrr}
\tablecaption{$V/V_m$ for galaxies and
quasars in the MS4, SMS4, and 3CRR samples.\label{tab13.vvm}}
\tablehead{
\colhead{} &
\multicolumn{3}{c}{MS4} & \colhead{} &
\multicolumn{3}{c}{SMS4} & \colhead{} &
\multicolumn{3}{c}{3CRR} \\
\cline{2-4} \cline{6-8} \cline{10-12} \\
\colhead{ID} &
\colhead{Mean} & \colhead{S.E.} & \colhead{$n$} & \colhead {} &
\colhead{Mean} & \colhead{S.E.} & \colhead{$n$} & \colhead {} &
\colhead{Mean} & \colhead{S.E.} & \colhead{$n$}}
\startdata
g & 0.60 & 0.02 & 176 & & 0.61 & 0.03 & 109 & & 0.63 & 0.02 & 129 \\
Q & 0.70 & 0.03 &  45 & & 0.76 & 0.04 &  21 & & 0.70 & 0.03 &  43 \\
\enddata
\end{deluxetable}

\clearpage

\begin{deluxetable}{rrlcrlrr}
\tablecaption{Comparison of quasar fractions and median linear sizes
$l_{med}$ of radio galaxies and quasars in the redshift range $0.5 < z
< 1$ from the MS4, SMS4, and 3CRR samples.  $f_Q$ stands for the
fraction of quasars, and ``ratio of $l_{med}$" for the median linear
size of quasars divided by the median linear size of galaxies.
\label{tab14.qf}}
\tablehead{
\colhead{} & \multicolumn{2}{c}{Galaxies} & \colhead{} &
\multicolumn{2}{c}{Quasars} & \colhead{} & \colhead{ratio} \\
\cline{2-3} \cline{5-6}
\colhead{} &
\colhead{} & \colhead{$l_{\rm med}$} & \colhead{} &
\colhead{} & \colhead{$l_{\rm med}$} &
\colhead{} & \colhead{of} \\
\colhead{Sample} &
\colhead{$n$} & \colhead{(kpc)} & \colhead{} &
\colhead{$n$} & \colhead{(kpc)} &
\colhead{$f_Q$} & \colhead{$l_{\rm med}$}}
\startdata
 & & & & & & & \\
MS4 & 39 & \merr{199}{-40}{+30} & & 11 & \merr{119}{-46}{+41} &
  $0.22 \pm 0.10$ & \merr{0.60}{-0.44}{+0.38} \\
 & & & & & & & \\
SMS4 & 25 & \merr{206}{-46}{+23} & & 7 & \merr{149}{-76}{+163} &
  $0.22 \pm 0.12$ & \merr{0.72}{-0.56}{+1.10} \\
 & & & & & & & \\
3CRR & 28 & \merr{192}{-90}{+94} & & 16 & \merr{66}{-24}{+30} &
  $0.36 \pm 0.15$ & \merr{0.34}{-0.59}{+0.67} \\
 & & & & & & & \\
\enddata
\end{deluxetable}

\clearpage

\begin{deluxetable}{crclrr}
\tablecaption{Median linear sizes for quasars and radio galaxies for
four different samples selected at low frequency.
The data for the Molonglo Quasar Sample (MQS) are
from \protect\citet{mqs.id}.\label{tab15.lin}}
\tablehead{
\multicolumn{3}{c}{} & \colhead{} & \colhead{} & \colhead{$l_{\rm med}$} \\
\cline{1-3}
\multicolumn{3}{c}{Sample} & \colhead{} & \colhead{$n$} & \colhead{(kpc)}}
\startdata
MQS & $S_{\rm 408} > 0.95$\,Jy & Q & & 69 & \merr{162}{-28}{+48} \\
 & & & & & \\
MS4 & $S_{\rm 408} > 4$\,Jy & Q & & 26 & \merr{116}{-43}{+44} \\
 & & & & & \\
SMS4 & $S_{\rm 178} > 10.9$\,Jy & Q & & 14 & \merr{154}{-37}{+109} \\
 & & & & & \\
3CRR & $S_{\rm 178} > 10.9$\,Jy & Q & & 37 & \merr{89}{-17}{+19} \\
 & & & & & \\
MS4 & $S_{\rm 408} > 4$\,Jy & g & & 103 & \merr{209}{-46}{+25} \\
 & & & & & \\
SMS4 & $S_{\rm 178} > 10.9$\,Jy & g & & 69 & \merr{214}{-22}{+31} \\
 & & & & & \\
3CRR & $S_{\rm 178} > 10.9$\,Jy & g & & 79 & \merr{234}{-32}{+58} \\
\enddata
\end{deluxetable}

\clearpage

\end{document}